\documentclass[runningheads,a4paper]{llncs}
\usepackage{booktabs}
\usepackage{booktabs, makecell, longtable}

\usepackage{rotating}
\usepackage{makecell}
\usepackage[flushleft]{threeparttable}
\usepackage{amssymb}
\usepackage{graphicx}
\usepackage{amsmath}
\usepackage{epstopdf}
\usepackage[hang]{subfigure}
\usepackage{bm}
\usepackage[utf8x]{inputenc} 
\usepackage{multicol, multirow}
\usepackage{url}
\urldef{\mailsa}\path|{silee, sjyoo}@sejong.ac.kr| 
\newcommand{\keywords}[1]{\par\addvspace\baselineskip
\noindent\keywordname\enspace\ignorespaces#1}
\newcommand{\RNum}[1]{\uppercase\expandafter{\romannumeral #1\relax}}

\begin{document}

\mainmatter  

\title{Multimodal Deep Learning for Finance: Integrating and Forecasting International Stock Markets}

\titlerunning{Multimodal Deep Learning for Finance}

%
%

\author{Sang Il Lee%
\and Seong Joon Yoo }
\authorrunning{Multimodal Deep Learning for Finance}

\institute{Department of Computer Engineering, Sejong University,\\
Seoul, Republic of Korea\\
\mailsa\\}

%
%

\toctitle{Lecture Notes in Computer Science}
\tocauthor{Authors' Instructions}
\maketitle

\begin{abstract}
In today's increasingly international economy, 
return and volatility spillover effects across international equity markets are 
major macroeconomic drivers
of stock dynamics. Thus, information regarding foreign markets 
is one of the most important factors in forecasting domestic stock prices. 
However, the cross-correlation between domestic and foreign markets is 
highly complex. Hence, it is extremely difficult to explicitly express 
this cross-correlation with a dynamical equation.
In this study, we develop stock return prediction models 
that can jointly consider 
international markets,
using multimodal deep learning. 
Our contributions are three-fold: (1) we visualize the transfer information
between South Korea and US stock markets by using scatter plots;
(2) we incorporate the information into the stock prediction models with the help of 
multimodal deep learning;
 (3) we conclusively demonstrate that the early and intermediate fusion models achieve a significant performance
boost in comparison with the late fusion and single modality models. Our study indicates
that jointly considering international stock markets can improve
the prediction accuracy and deep neural networks are highly effective for
such tasks.

\keywords{Stock prediction, deep neural networks, multimodal, data fusion, international stock markets}
\end{abstract}

\section{Introduction}
\label{intro}
\subsection{ Aims and scope of the study}
The interdependence between international stock markets has been steadily increasing in recent years.  
In particular, after the stock market crash of $1987$, the interdependence increased significantly \cite{eun89}, and more recently, this interdependence was widely noticed during the global financial crisis of $2007$ \cite{bek09}. Both originated in the US and resulted in a sharp decline 
in the stock prices of international stock markets,
rapidly spreading to other countries.
The crisis clearly confirmed that the financial events originating in one market are not isolated to that particular market but are also transmissible across international borders. Currently, this internationalization is a common phenomenon and expected to accelerate.

The goal of our study is to investigate the contribution of additional international market information in stock prediction by using deep learning.
Typically, this interconnection has not been considered in stock prediction unlike various data categories such as country-specific price, macroeconomic, news, and fundamental data.
We considered the South Korean and US stock markets with non-overlapping stock exchange trading hours as a case study and studied the one-day ahead stock return prediction of the South Korean stock market by combining the data of the two markets. The combination of the markets is particularly fascinating due to their different behaviors: the US markets have a long-run upward trend whereas the South Korean markets do not. Therefore, the possible existing correlations between them are not just the result of the continued global economic growth. 
We utilized the daily trading data (i.e., opening, high, low, and closing prices) of both markets, which is publicly available and quantifies the daily movement of the markets. The publicity ensures that our results are more likely to be independent and easily integrated with other data, serving as a prototypical model.
 
We designed multimodal deep learning models to extract cross-market correlations by concatenating features at early, intermediate, and late fusions between modalities. The models places a different emphasis on intra and inter market correlations depending on the markets to be tested. 
The experiments showed that the early and intermediate fusions
achieve better prediction accuracy than the single modal prediction and late fusions. 
This indicates that multimodal deep learning can capture cross-correlations from stock prices despite their low signal-to-noise ratio. It also indicates that, when optimizing prediction models, cross-market learning provides opportunities to improve the accuracy of stock prediction, even when the shared trends in markets are scarce. 

The remainder of this paper is organized as follows.
Section \ref{subsec:1.2} discusses the connections to existing work.
Section \ref{sec:2} introduces the
US and South Korean (KR) international stock markets.
Section \ref{sec:3} discusses data and preprocessing methods.
Section \ref{sec:4} describes a basic architecture 
for deep neural networks and illustrates three prediction models.
Section \ref{sec:5} presents information on the training of the deep neural networks.
Section \ref{sec:6} presents
the prediction accuracy of the models and discusses their capacity.
Finally, Section \ref{sec:7} presents the concluding remarks and future scope
of the study.

\subsection{Connections with previous studies}
\label{subsec:1.2}
Over the past few decades, machine learning techniques, such as artificial neural networks (ANNs), genetic algorithms (GAs), support vector machines (SVM),
and natural language processing (NLP), have been widely employed to model financial data. For example, a genetic classifier designed to control the activation of ANNs \cite{arm05}, the genetic algorithms approach to feature discretization in ANNs \cite{kim00},  the wavelet de-noising-based ANN  \cite{wan11}, wavelet-based ANN \cite{saa14}, and the surveys on sentiment analysis \cite{xin18} and machine learning \cite{ats09,cav16}.

Machine learning techniques help to mitigate the difficulties in modeling, 
such as the existence of nonlinear behaviors 
in financial variables, the non-stationarity of 
relationships among the relevant variables, and a low signal-to-noise ratio.
In particular, deep learning is becoming a
promising technique for modeling financial complexity, owing to
its ability to extract relevant information 
in complex, real-world world data \cite{ben13}. For example,
stock prediction based on long short-term memory (LSTM) networks \cite{fis17},
deep portfolios based on deep autoencoders \cite{hea17},
threshold-based by using recurrent neural networks \cite{lee18}, and
deep factor models involving deep feed-forward networks \cite{nak18}, and LSTM networks
\cite{nak19}.

A major challenge for further research in this area is the simultaneous consideration of the numerous factors in financial data modeling.
In the search for factors that explain the cross-sectional 
expected stock returns, numerous potential candidates have been found by using econometric models. For example, accounting data, macroeconomic
data, and news \cite{coc11,har15,mcl16,hou17,fen19}.
Stock price predictions that consider a few pre-specified factors 
may lead to incorrect forecasting as they reflect partial information 
or an inefficient combination of the factors.
Thus, currently, one of the most important tasks in finance
is to develop a method that effectively integrates diverse factors 
in prediction processes. 

A few recent studies have begun to combine
financial data using deep learning.
Xing et al. \cite{xin18-2} dealt with
the price, volume, and sentiment data to build
a portfolio using LSTM networks. Bao et al. \cite{bao17} used
trading data (prices and volume),
technical indicators, and macroeconomic data 
(exchange and interest rates) to predict stock prices
by combining the wavelet transforms (WT), stacked autoencoders (SAEs), and LSTMs. 
The fusion strategy of these studies 
concatenates the data into
the input layers, known as
an early fusion. However, because hidden layers in 
such approaches are exposed to cross-modality information, it could be more difficult to 
use them specifically to extract the essential 
intra-modality relations during
training.
In this study, to effectively integrate
financial data,
we introduce a systematic fusion approach, i.e.,
early, intermediate, and late fusions, by considering the international stock markets as a case study.

International market dynamics
has been a controversial issue in 
financial academia and industries due to the 
increasing economic globalization.
Although stock market integration 
is intuitively obvious in an era of free trade and globalization, 
the underlying mechanisms are highly complex and not easily understood.  
Financial economists have developed models for describing
The dynamic interdependency among major world stock exchanges using
econometric tools such as vector autoregression (VAR) and
autoregressive conditional heteroskedastic (ARCH) models
\cite{cam92,kar96}. 
They have attempted to find underlying reasons behind the 
interdependence, providing 
possible scenarios of mechanisms in terms of 
deregulation \cite{tay89,jeo91}, international business cycles \cite{kas92}, regional affiliations and trade linkages \cite{bac96}, and regional economic integration \cite{boo97}. 
However, despite the advantage of such approaches in explaining the underlying mechanism,
they generally only deal with a small number of financial variables; 
and as a result describe only a partial aspect of the complex financial reality,
which is actually characterized by multidimensional and nonlinear characteristics.
Thus, international markets are a good
case study for the effectiveness of 
deep learning in financial data fusion. Furthermore, modeling international markets is important in practice because
investors and portfolio managers need to continually assess international
information and adjust their portfolios accordingly, in order to take the benefits of portfolio diversification \cite{syr04}.

Technically,
we were inspired by the success of the multimodal deep learning technique \cite{ng11,zh15,sri12} 
in computer science. 
The main advantage of deep learning is the ability to automatically learn
hierarchical representations from raw data, which can then be extended to cross-modality shared
representations at different levels of abstraction \cite{ng11,sri12}.
Multimodal deep learning has been widely applied 
to multiple channels of communication, such as
auditory (words, prosody, dialogue acts, and rhetorical structure)
and visual (gesture, posture, and graphics), achieving 
better prediction accuracy than 
approaches using only single-modality data.

\section{International stock markets: US and KR}
\label{sec:2}

We consider two international stock markets of South Korea (KR) and the US.
They are effective cases for studying the spillover effect 
because the trading time horizons of these markets do not overlap. 
The US stock market opens at 9:30 a.m. and closes at 4:00 p.m. (EST time), whereas
the KR stock exchange opens at 9:00 a.m. and closes at 3:30 p.m. (KST time). 
The KR market opens three hours after the US market closes. 
Due to the non-overlapping time zones, 
the closing prices of the US market index affect 
the opening prices of the KR market index, and vice versa.

There is significant empirical evidence on the correlative behavior between the two markets.
Na and Sohn \cite{na11}
investigated the co-movement between
the Korea composite stock index (KOSPI) and 
the world stock market indexes using association rules. 
They found that
the KOSPI tends to move in the same direction as the stock market indices in the US and Europe, and in the opposite direction to
those in other East Asian counties, including both Hong Kong and Japan, which have competitive relationships with KR.
Jeon and Jang \cite{jeo04} found that the US market plays
a leading role in the KR stock market by applying the vector 
autoregression (VAR) model to the daily stock prices in both nations.
Lee \cite{lee06} statistically showed a significant volatility spillover effect
between them.  

\begin{figure}[t]
\centering
     \scalebox{0.5}
  {
	\includegraphics{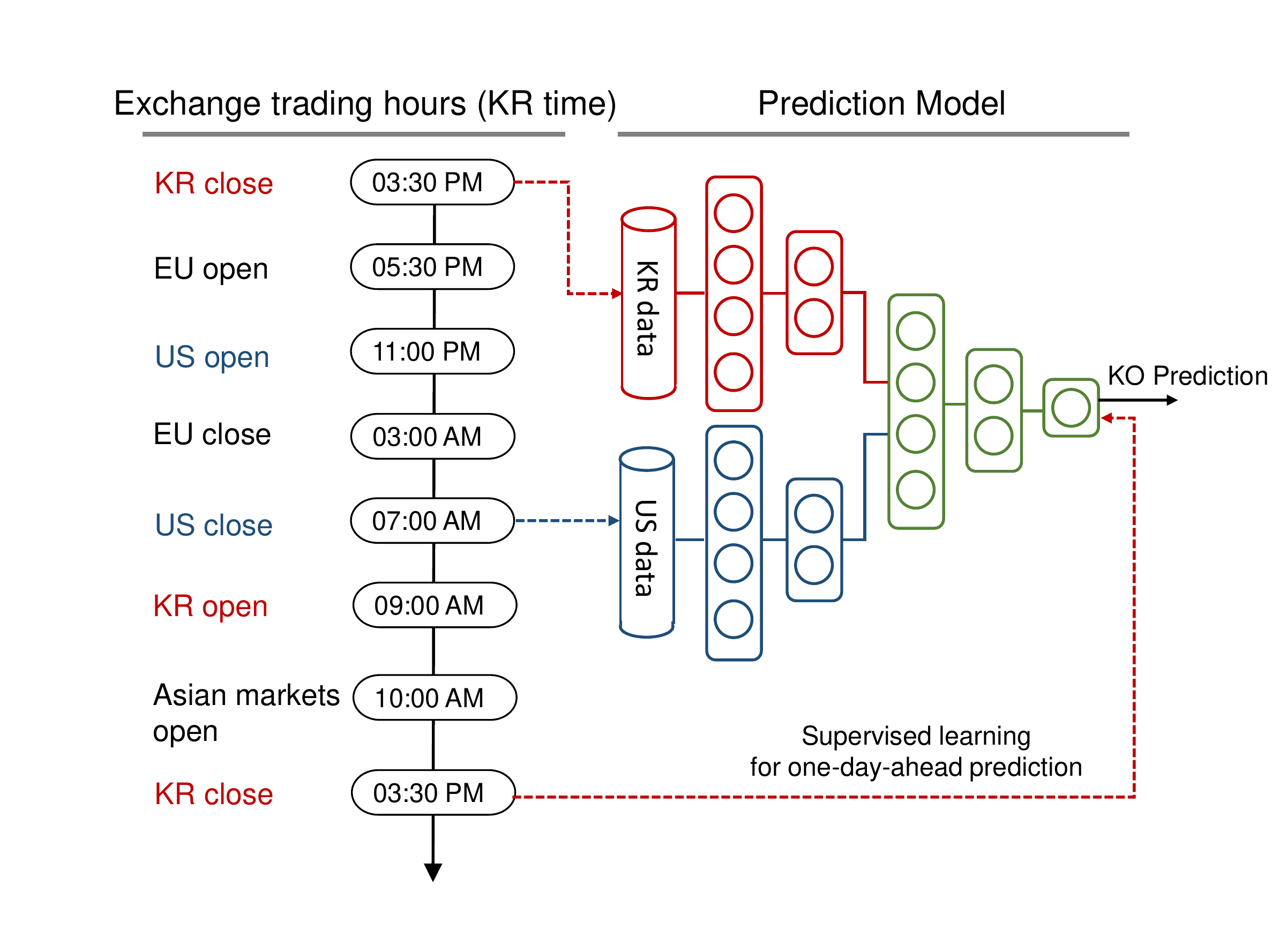}

   }
\caption{Schematic diagram
of the model integrating KR and US stock market prices and 
predicting KR market prices.
}
\label{DNN_timezone}
\end{figure}

Overall, the results of previous studies based on traditional financial models and primarily linear regression models, have consistently demonstrated the existence of an interrelationship between the two markets by identifying statistically significant explanatory variables. The objective of this study is to capture
this interrelationship by using multimodal deep learning and utilize it as complementary information for stock prediction.
Figure $\ref{DNN_timezone}$ shows a schematic diagram
of the model that integrates the KR and US stock market prices and predicts the KR market (the neural network will be discussed in detail in Section \ref{sec:4}). 

\section{Data and preprocessing}
\label{sec:3}
\subsection{International Market Indexes}
We used the KOSPI (KO) index 
as a proxy of the KR stock markets and the Standard and Poor's 500 (SP), NASDAQ (NA), and Dow Jones Industrial Average (DJ) indexes as proxies of the US stock market. The KO is a highly representative index of the KR stock markets as it tracks the performance of all common shares listed on the KR stock exchange, based on capitalization-weighted schemes.
The DJ is a price-weighted index composed of 30 large industrial stocks. The SP is a value-weighted index of 500 leading companies in diverse industries of the US economy. The SP index covers 80$\%$ of the value of US equities and therefore provides an aggregate view of overnight information in the United States. The NA is weighted by capitalization of the stocks included in its index and contains
stocks in large technology firms, such as Cisco, Microsoft, and Intel.

\subsection{Raw data} 
\begin{figure}[t]
\centering
  \scalebox{0.4}
  {
	\includegraphics{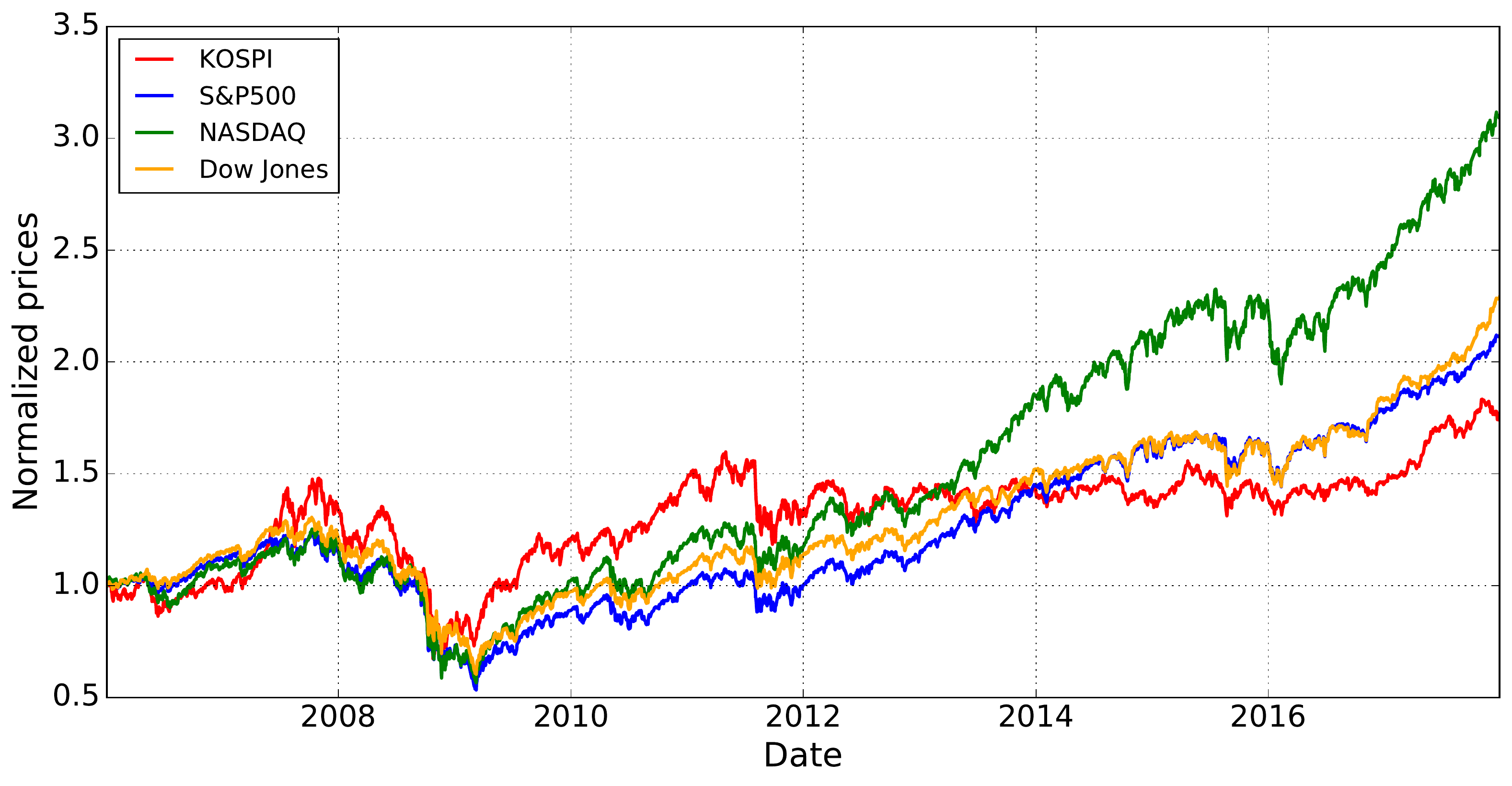}
		   }
\caption{Normalized KOSPI, S$\&$P500, DAIJ, and NASDAQ
indexes over the period from 2006 to 2017 obtained by 
subtracting the mean from each original value and dividing by the standard
deviation.}
\label{fig_stock_price}
\end{figure}

The daily market data for the four indexes are obtained from Yahoo Finance and contain daily trading data, such as opening prices (Open), high prices (High), low prices (Low), adjusted closing prices (Close), and end-of-day volumes. The data are from the period between January 1st, 2006 and December 31st, 2017 (Fig. \ref{fig_stock_price}). The data from days where either one of the stock markets
was closed was excluded from our data set.

\subsection{Training, validation, and test set}
All data are divided into a training dataset (70$\%$) for developing
the prediction models 
and test set (30$\%$) for evaluating its predictive ability.  
30$\%$ of the training set was used as a validation set.

\subsection{Feature construction}
We seek predictors in order to predict the daily (close-to-close) 
KO return at time $t+1$ $r_{t+1}$, 
given the feature vector ${\boldsymbol x}_{t}$ extracted
from the trading data available at time $t$.  
To describe the movement of 
the indexes, we defined a set of meaningful features at time $t$ as follows:
\begin{enumerate}
\item  $\displaystyle \textrm{Daytime},\ \ \textrm{High-to-Close Return}:=\textrm{DHTC}^{\cdot}_{t}=\frac{\textrm{Hight}_{t}-\textrm{Close}_{t}}{\textrm{Close}_{t}}$,\\
\item $\displaystyle \textrm{Daytime},\ \ \textrm{Open-to-Close return}:=\textrm{DOTC}^{\cdot}_{t}=\frac{\textrm{Open}_{t}-\textrm{Close}_{t}}{\textrm{Close}_{t}}$,\\
\item $\displaystyle \textrm{Daytime},\ \ \textrm{Low-to-Close return}\ \ :=\textrm{DLTC}^{\cdot}_{t}=
\frac{\textrm{Low}_{t}-\textrm{Close}_{t}}{\textrm{Close}_{t}}$,\\
\item $\displaystyle \textrm{Overnight}, \textrm{Close-to-Close return}:=\textrm{OCTC}^{\cdot}_{t}=
\frac{\textrm{Close}_{t}-\textrm{Close}_{t-1}}{\textrm{Close}_{t-1}}$, \\
\item $\displaystyle \textrm{Overnight}, \textrm{Open-to-Close return}:=\textrm{OOTC}^{\cdot}_{t}=
\frac{\textrm{Open}_{t}-\textrm{Close}_{t-1}}{\textrm{Close}_{t-1}}$.
\end{enumerate}
The features describe the daily movement of
stock indexes: $\textrm{DHTC}^{\cdot}_{t}$ for the highest daytime movement, 
$\textrm{DLTC}^{\cdot}_{t}$ for the lowest daytime movement,
$\textrm{DOTC}^{\cdot}_{t}$ for the daytime movement, 
$\textrm{OOTC}^{\cdot}_{t}$ for the opening jump 
responding to the overnight information, and
$\textrm{OCTC}^{\cdot}_{t}$ for the total movement 
reflecting all information available
at time $t$. 

Let us denote the feature vector for each modality as  $
{\boldsymbol x}^{i}_{t}=[\textrm{DHTC}_{t}^{i}, \textrm{DOTO}_{t}^{i}, 
\textrm{DLTC}_{t}^{i}, \textrm{OCTC}_{t}^{i}, \textrm{OOTC}_{t}^{i}]^{T}
$, where $i \in \{\textrm{KO, SP, DJ, NAS}\}$, and
$\textrm{US}\in \{\textrm{SP,NA,DJ}\}$. An input feature ${ \boldsymbol x}_{t}$ for multimodal models at time $t$ is 
the combination of ${\boldsymbol x}_{t}^{\textrm{KO}}$ and ${\boldsymbol x}_{t}^{\textrm{US}} $, 
depending on the multimodal deep learning architecture.
 Note that we did not include returns across markets, 
 such as SP Close-to-KO Close return=$\textrm{Close}_t^{\textrm{SP}}/\textrm{Close}_t^{\textrm{KR}}-1$, because they are statistically non-stationary at any conventional significance level. In the following, we will use the notation 
$\textrm{OCTC}^{\textrm{KO}}_{t}$ and $r_{t}$ interchangeably to denote the daily 
close-to-close return on the KO index.

To improve the accuracy of the prediction and prevent complications 
arising from convergence during training, we normalized the individual feature 
into the range $[\min,\max]$, using the following formula:
\begin{align}
x \longleftarrow  \frac{x-\min_{\textrm{train}}}{\textrm{max}_{\textrm{train}}-\textrm{min}_{\textrm{train}}}
(\textrm{max}-\min)+\min,
\label{min-max}
\end{align}
where $x$ on the right side represents the normalized value of data $x$ on the right side;  
$\max_{\textrm{train}}$ and $\min_{\textrm{train}}$ denote the maximum and minimum values of data $x$, respectively; these were estimated using only the training set to avoid look-ahead biases and then applied to the validation and test sets.

\subsection{Association of the two markets}
\begin{figure}[h]
  \centering
  \subfigure[$\beta_{0}=-0.000$, $\beta_{1}=0.164$]{%
    \includegraphics[width=0.47\linewidth]{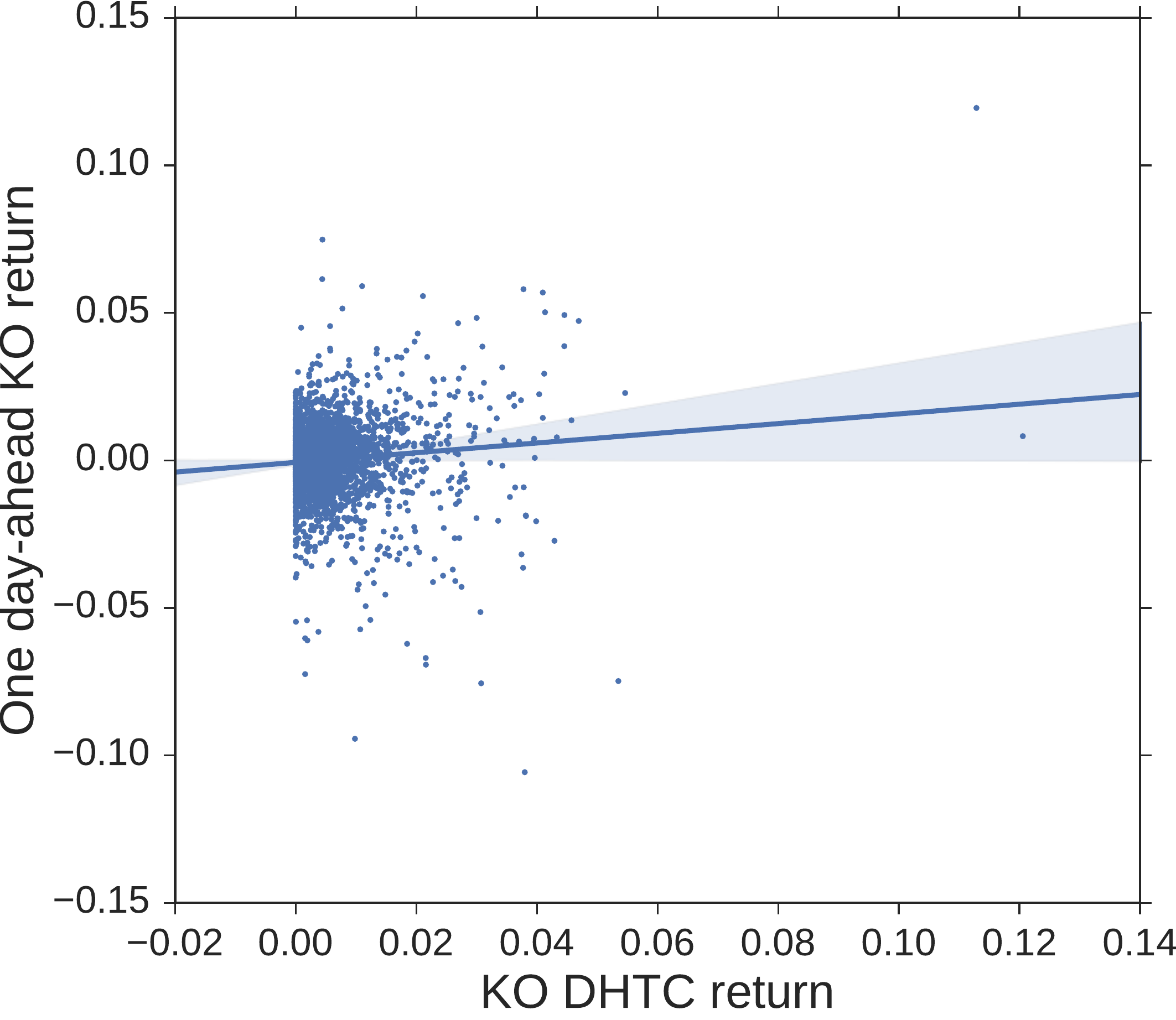}%
  }
  \subfigure[$\beta_{0}=-0.000$, $\beta_{1}=0.087$]{%
    \includegraphics[width=0.47\linewidth]{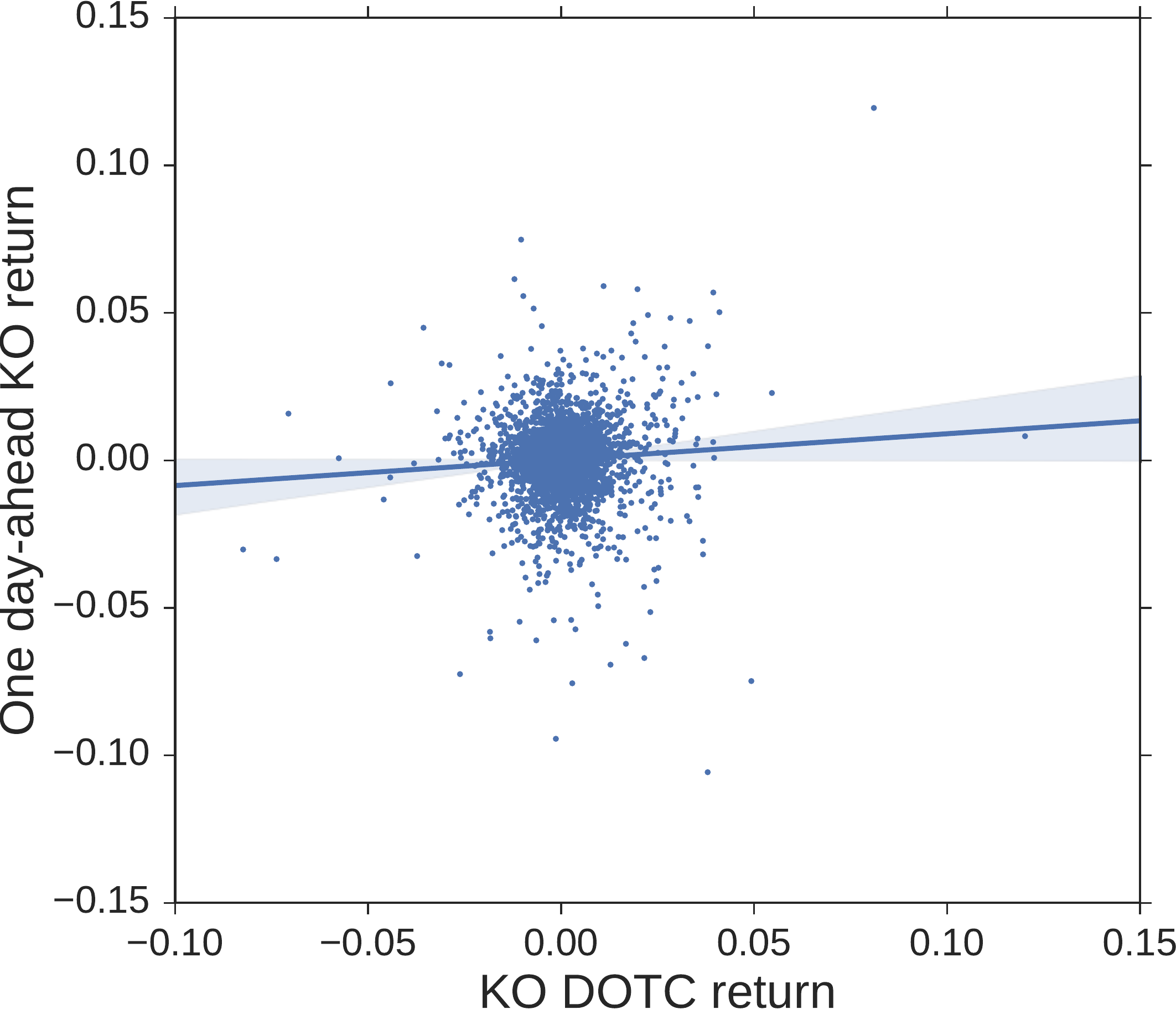}%
  }\\
    \subfigure[$\beta_{0}=-0.000$, $\beta_{1}=0.057$]{%
    \includegraphics[width=0.47\linewidth]{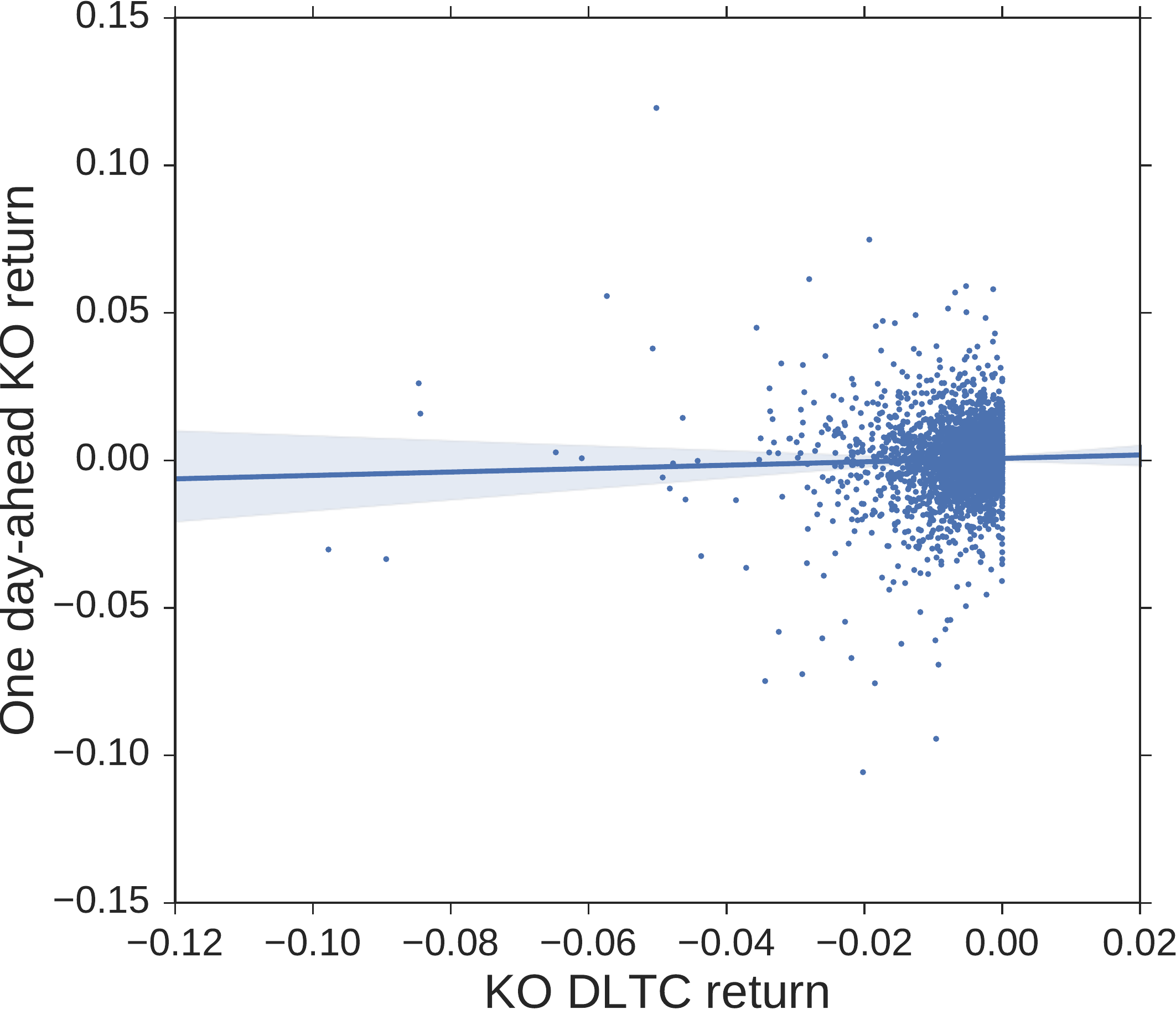}%
  }
  \subfigure[$\beta_{0}=-0.000$, $\beta_{1}=0.002$]{%
    \includegraphics[width=0.47\linewidth]{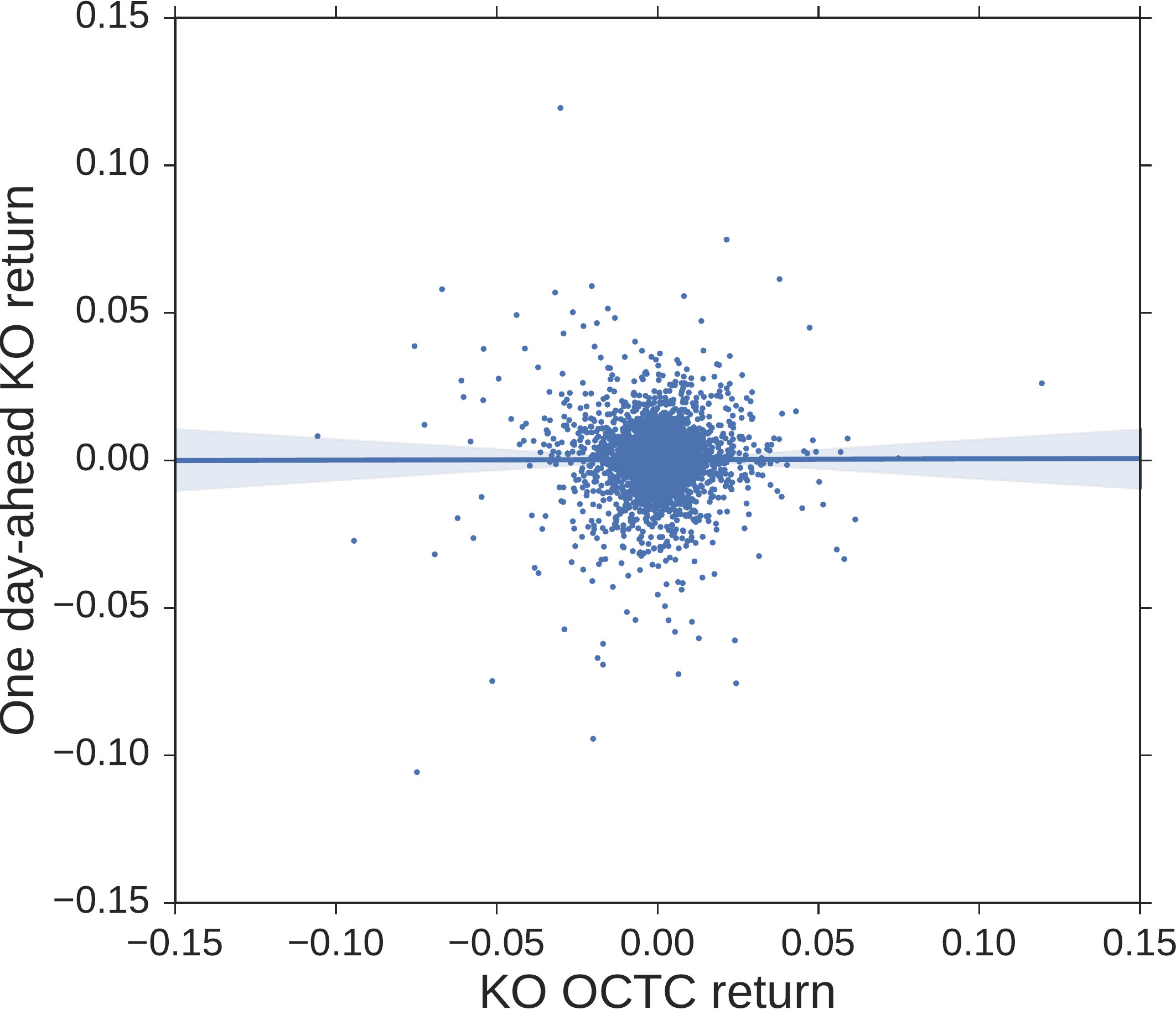}%
  }\\
    \subfigure[$\beta_{0}=-0.000$, $\beta_{1}=0.112$]{%
    \includegraphics[width=0.47\linewidth]{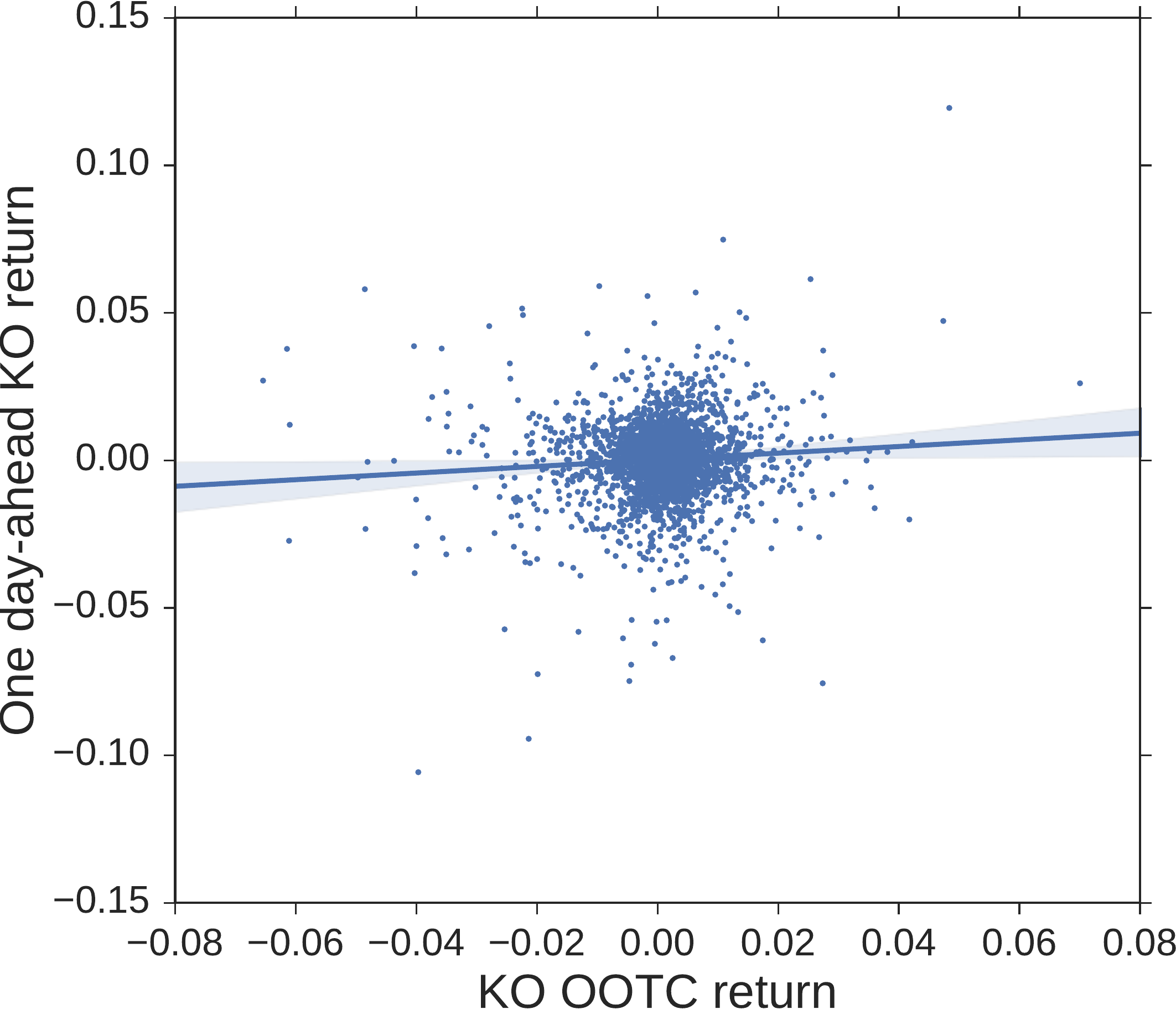}%
  }
  \caption{Scatter plots of the pairs (KO feature, one-day-ahead KO return)   from January 1st, 2006 to December 31st, 2017, 
with a regression line and associated 95$\%$ bootstrapped confidence intervals. The regression equation is given by $\textrm{DOTC}_{t+1}=\beta_{0}+\beta_{1}x^{\textrm{KR}}_{t}+\varepsilon$, where $\beta_{0}$, $\beta_{1}$, and $\varepsilon$ are 
the intercept, slope, and random disturbance, respectively.}
  \label{fig:scatter_KO}
\end{figure}

\begin{figure}[h]
  \centering
  \subfigure[$\beta_{0}=0.002$, $\beta_{1}=-0.453$]{%
    \includegraphics[width=0.47\linewidth]{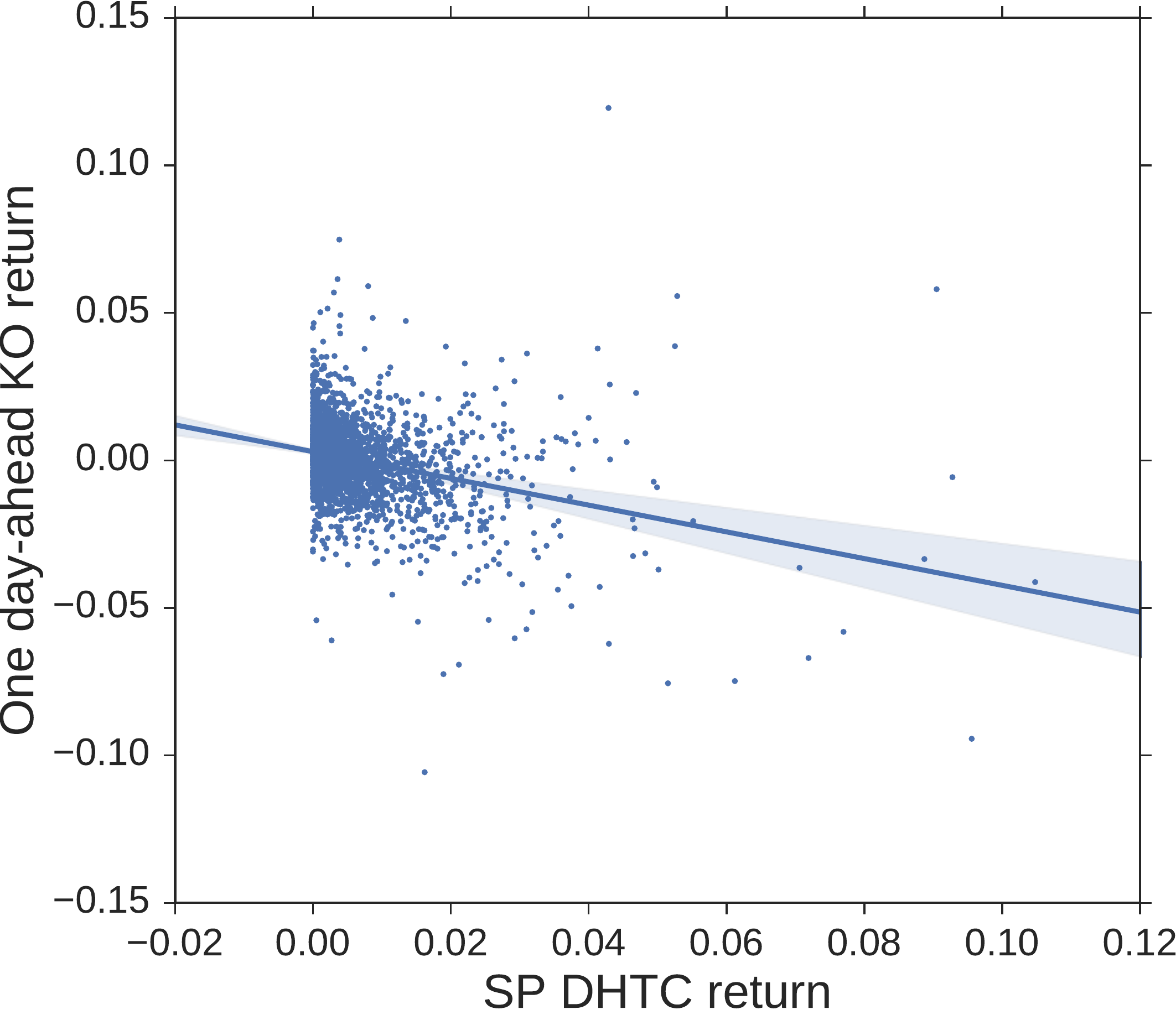}%
  }
  \subfigure[$\beta_{0}=0.000$, $\beta_{1}=-0.431$]{%
    \includegraphics[width=0.47\linewidth]{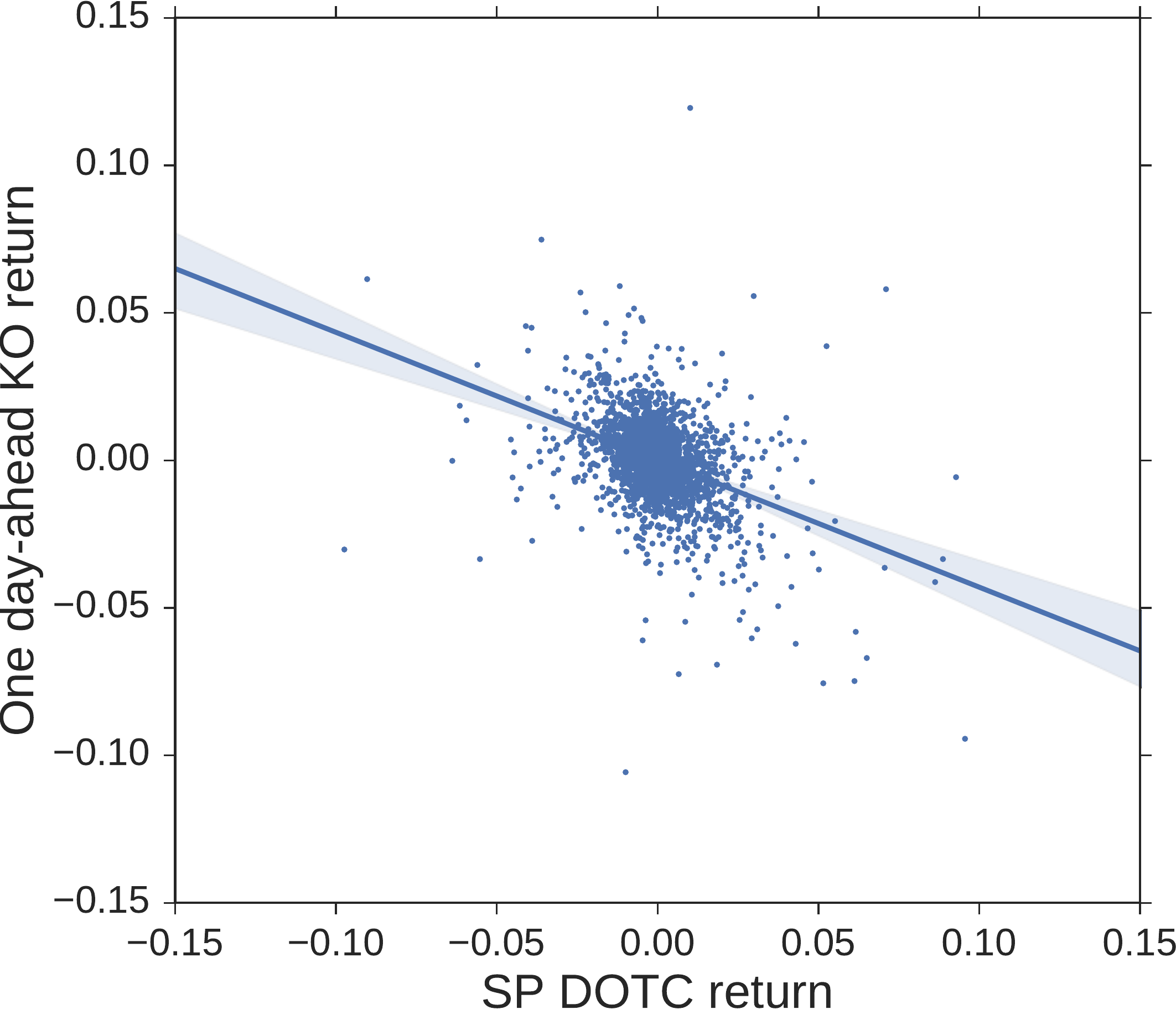}%
  }\\
    \subfigure[$\beta_{0}=-0.001$, $\beta_{1}=-0.332$]{%
    \includegraphics[width=0.47\linewidth]{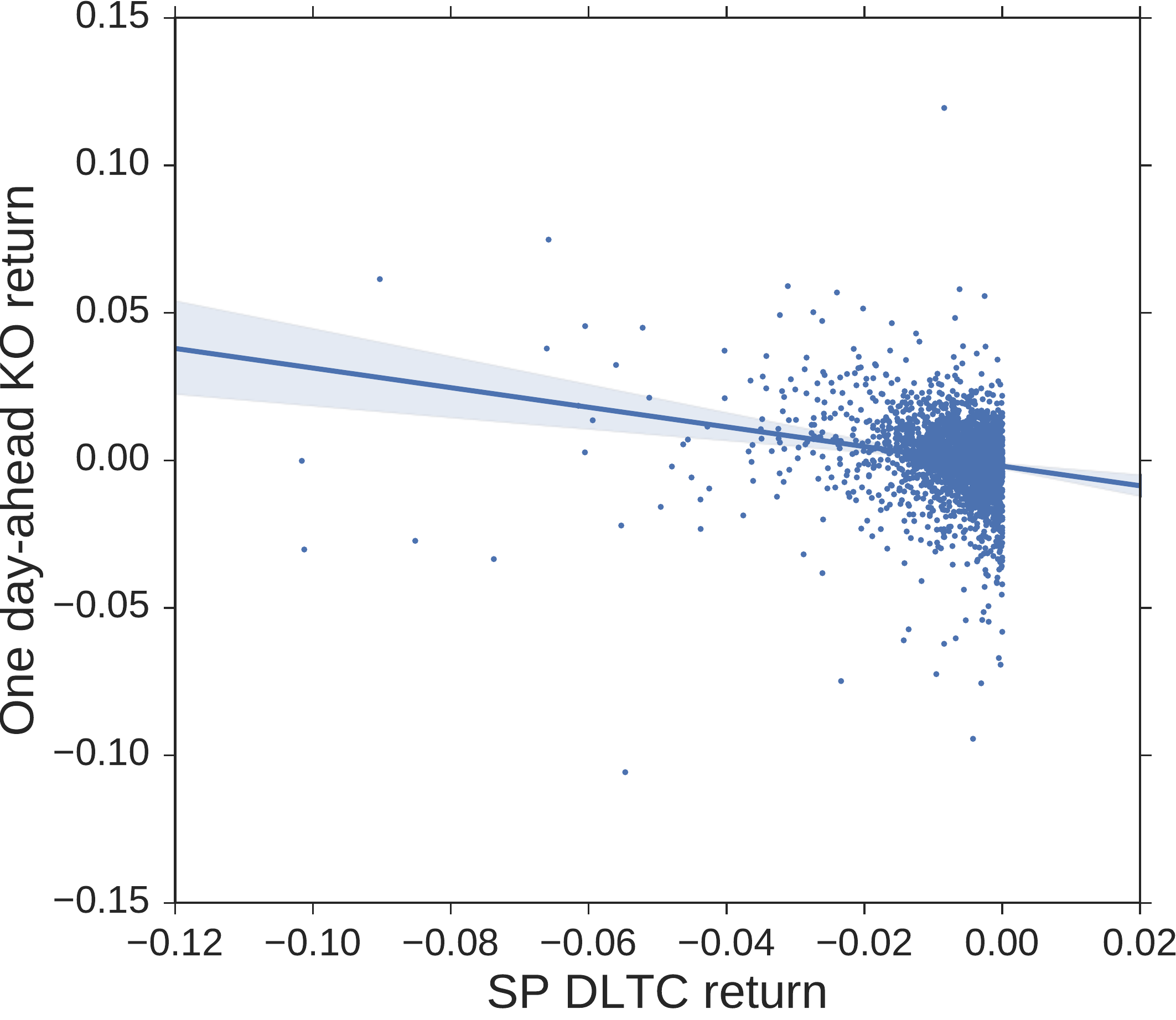}%
  }
  \subfigure[$\beta_{0}=0.000$, $\beta_{1}=0.385$]{%
    \includegraphics[width=0.47\linewidth]{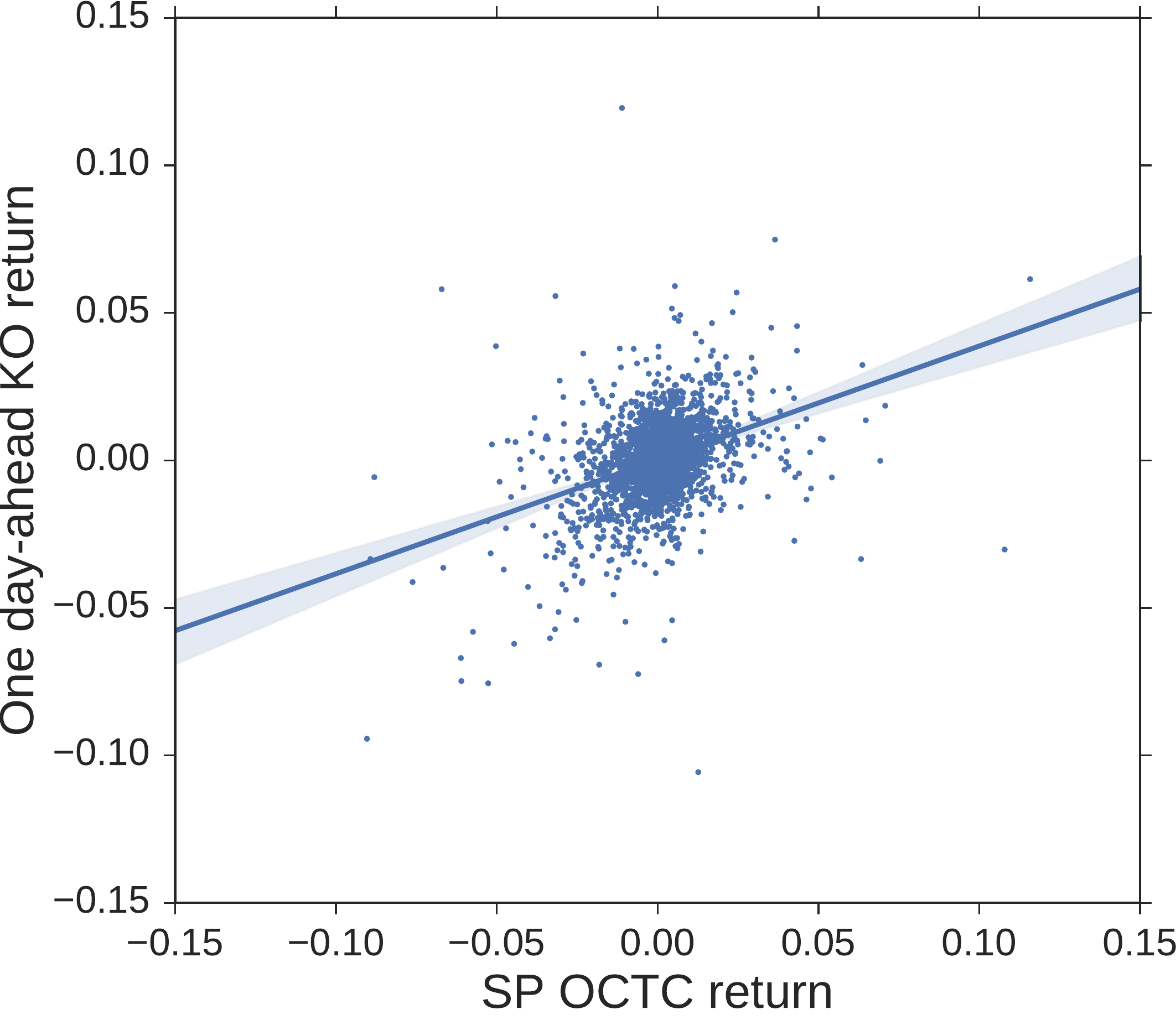}%
  }\\
    \subfigure[$\beta_{0}=0.000$, $\beta_{1}=0.302$]{%
    \includegraphics[width=0.47\linewidth]{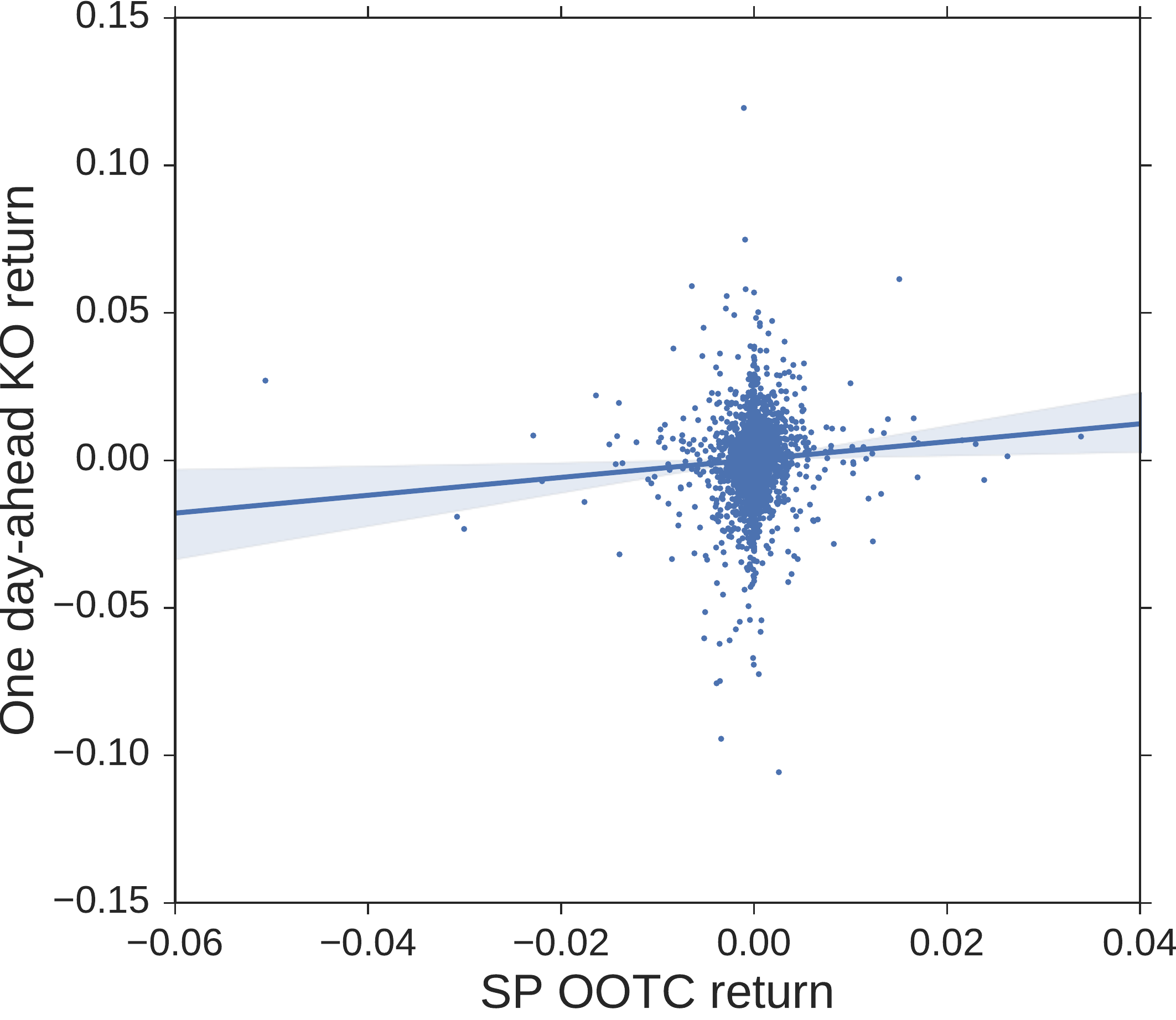}%
  }
  \caption{Scatter plots of the pairs (SP feature, one-day-ahead KO return)
from January 1st, 2006 to December 31st, 2017, 
with a regression line and associated 95$\%$ confidence interval. The regression equation is given by $\textrm{DOTC}_{t+1}=\beta_{0}+\beta_{1}x^{\textrm{SP}}_{t}+\varepsilon$, where $\beta_{0}$, $\beta_{1}$, and $\varepsilon$ are 
the intercept, slope, and random disturbance, respectively.}
  \label{fig:scatter_SP}
\end{figure}

For an intuitive understanding, we visualize the patterns between the features and target by using 
scatter plots with a regression best-fit line.
 Figure \ref{fig:scatter_KO} shows the scatter plots 
for the pairs of KO features and the one-day-ahead returns. There is an extremely weak positive linear association, which is described by the shallow slopes of the
regression lines from 0.002 to 0.164 and significant variation around the linear regression lines.
As shown in Fig. \ref{fig:scatter_SP},
the scatter plots for the SP features exhibit more diverse patterns, i.e., positive as well as negative slopes, with relatively steeper slopes from $-0.453$ to 0.385
 and significant variation around the linear regression lines. 
The steeper slopes exhibit a certain extent of a spillover effect 
from the US daytime stock market
to the next day KR stock market. This implies that the US and KR 
markets share a certain amount of information. We ultimately intend to capture this information by using multimodal deep learning.

\section{Multimodal deep learning network model}
\label{sec:4}
\subsection{Deep neural network}
The multimodal deep neural network consists of deep neural networks (DNNs), which are a sequence of fully connected layers.
The DNN can extract high-level features from raw data through statistical learning over a large amount of data to obtain an effective representation of input data.

Suppose that we are given a training data set $\{ {\boldsymbol x}_{t}\}^{T}_{t=1}$ and
a corresponding label set $\{ {r_{t}} \}_{t=2}^{T+1}$, where $T$ denotes 
the number of days in the period of the training set.
The DNN consists of an input layer $L_{0}$, an output 
layer $L_{out}$, and H hidden layers $L_{h} (h\in \{1,2,\ldots,H \})$ between the input and output layers. Each hidden layer $L_{h}$ is a set of
several units, which could be arranged as a vector ${\boldsymbol a}\in \mathbb{R}^{|L_{h}|}$,
where $|L_{g}|$ denotes the number of units in $L_{h}$.
The units in $L_{h}$ are recursively defined as a nonlinear transformation
of the $h-1$-th layer:
\begin{align}
{\boldsymbol a}_{h}=f({\boldsymbol W}_{h}^{\textrm{T}} {\boldsymbol a}_{h-1}+{\boldsymbol b}_{h}),
\end{align}
where the weight matrix ${\boldsymbol W}_{h}\in \mathbb{R}^{|L_{h-1}|\times |L_{h}|}$,
the bias vector ${\boldsymbol b}_{h}\in \mathbb{R}^{|L_{h}|}$, and $f(\cdot)$,
where the weight matrix ${\boldsymbol W}_{h}\in \mathbb{R}^{|L_{h-1}|\times |L_{h}|}$.
The nonlinear activation function $f(\cdot): \mathbb{R}^{N_{l}\times 1} \rightarrow
\mathbb{R}^{N_{l}\times 1}$ acts entry-wise on its argument and
the units ${\boldsymbol a}_{0}$ in the input layer $L_{0}$ are the feature vectors.
According to the daily return regression task, 
a single unit with a linear activation function in the output layer is
used in the output layer $L_{out}$. 
Then, given the input ${\boldsymbol a}_{0}={\boldsymbol x}_{t}$, the one day-ahead return prediction
$\hat{r}^{\cdot}_{t+1}$ is given by
\begin{align}
\hat{r}^{\cdot}_{t+1}=W_{out}^{\textrm{T}}{\boldsymbol a}_{H},
\end{align}
where 
$W_{out} \in \mathbb{R}^{|L_{H}|}$ and ${\boldsymbol a}_{H}$ is the unit in the final
hidden layer $L_{H}$.

\subsection{Single and multimodal deep networks for stock prediction}

\begin{figure}[t]
\centering
  \scalebox{1.}
  {
    \scalebox{0.45}
  {
  		\includegraphics{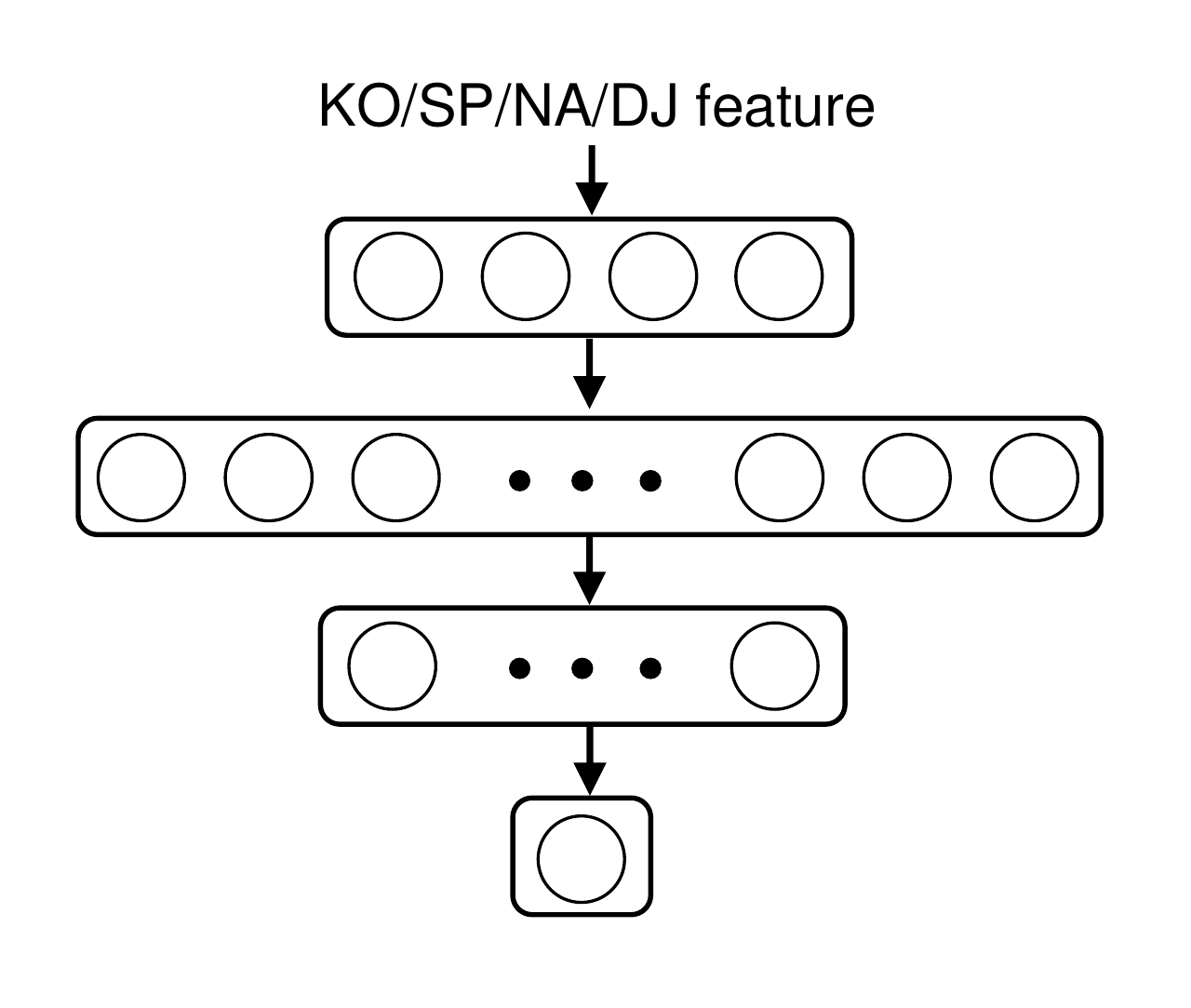}
  		}
   \scalebox{0.4}
  {
		\includegraphics{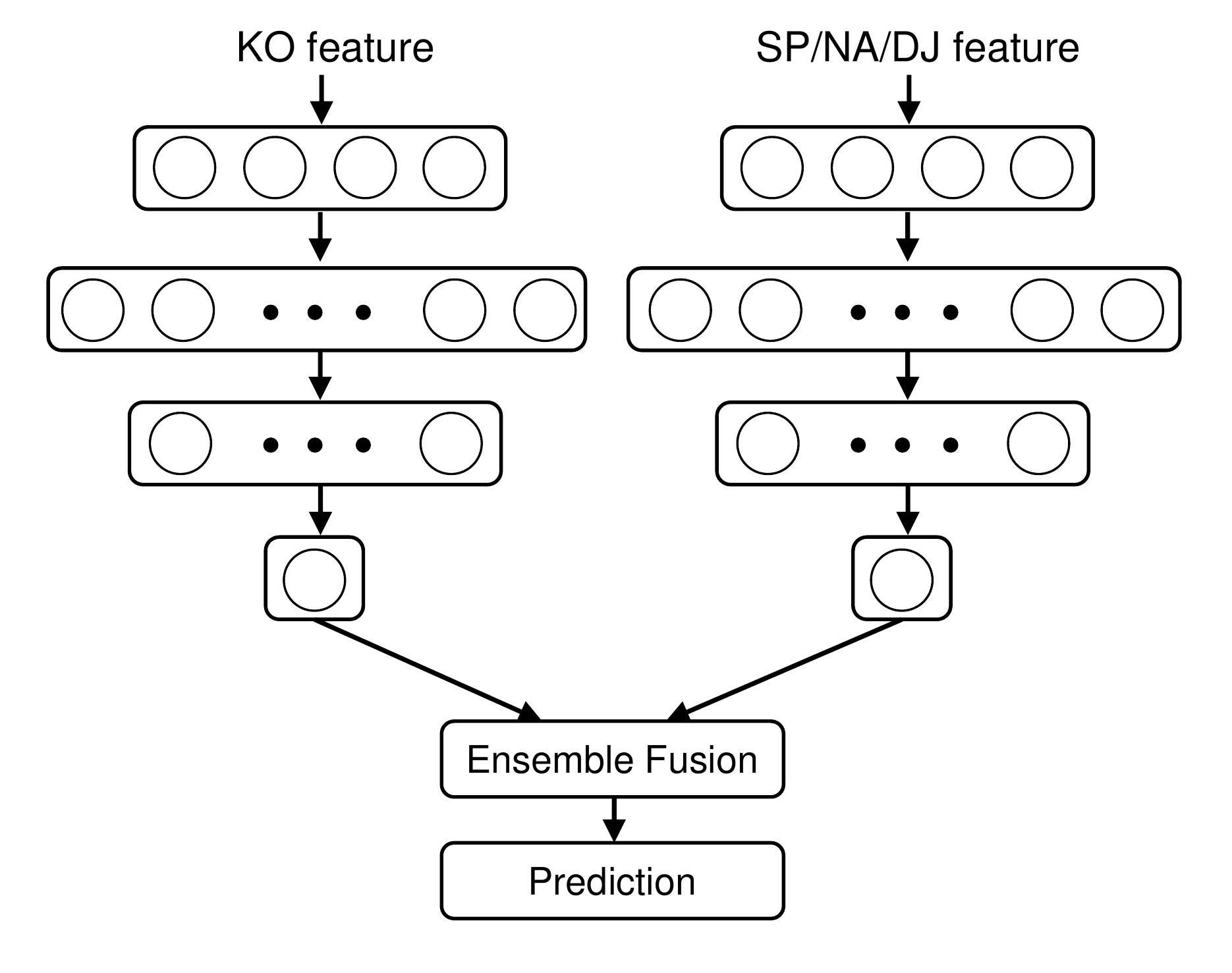}
}
   }
\caption{The KO/SP/NA/DJ-Only DNN model is shown in the left figure, where
the input feature is given by KO/SP/NA/DJ, respectively. 
The ensemble fusion model is shown in the right figure, where
$\{ \hat{y}^{\textrm{KO}}_{t+1}\}$ and $\{ \hat{y}^{\textrm{US}}_{t+1}\}$ are individually produced from each DNN, and
the final predictions $\{ \hat{y}_{t+1}\}$ are obtained using rules. 
}
\label{fig_DNN1}
\end{figure}

\begin{figure}[t]
\centering
   \scalebox{1.}
   {
   \scalebox{0.45}
  {
		\includegraphics{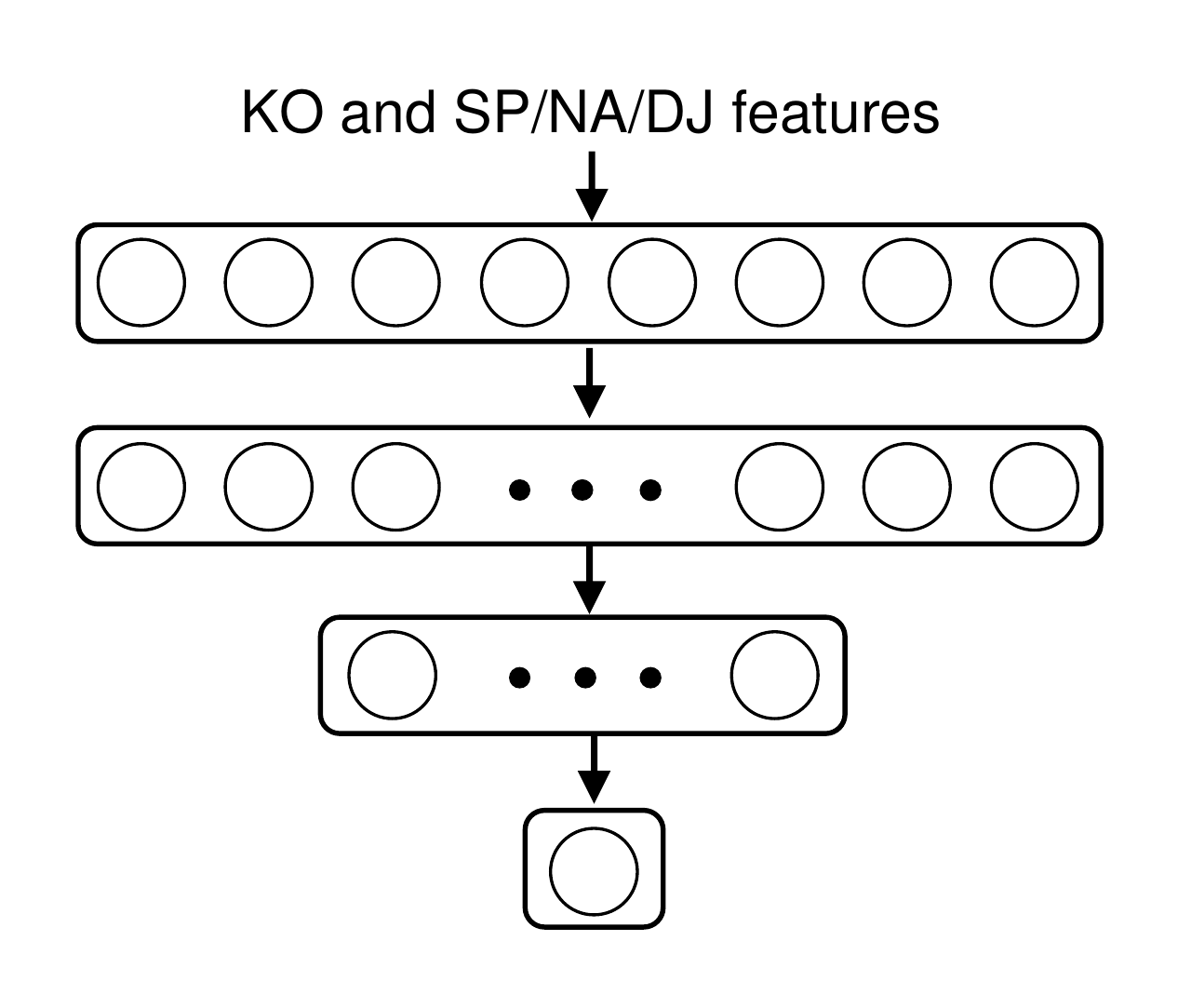}
}
     \scalebox{0.45}
  {
	\includegraphics{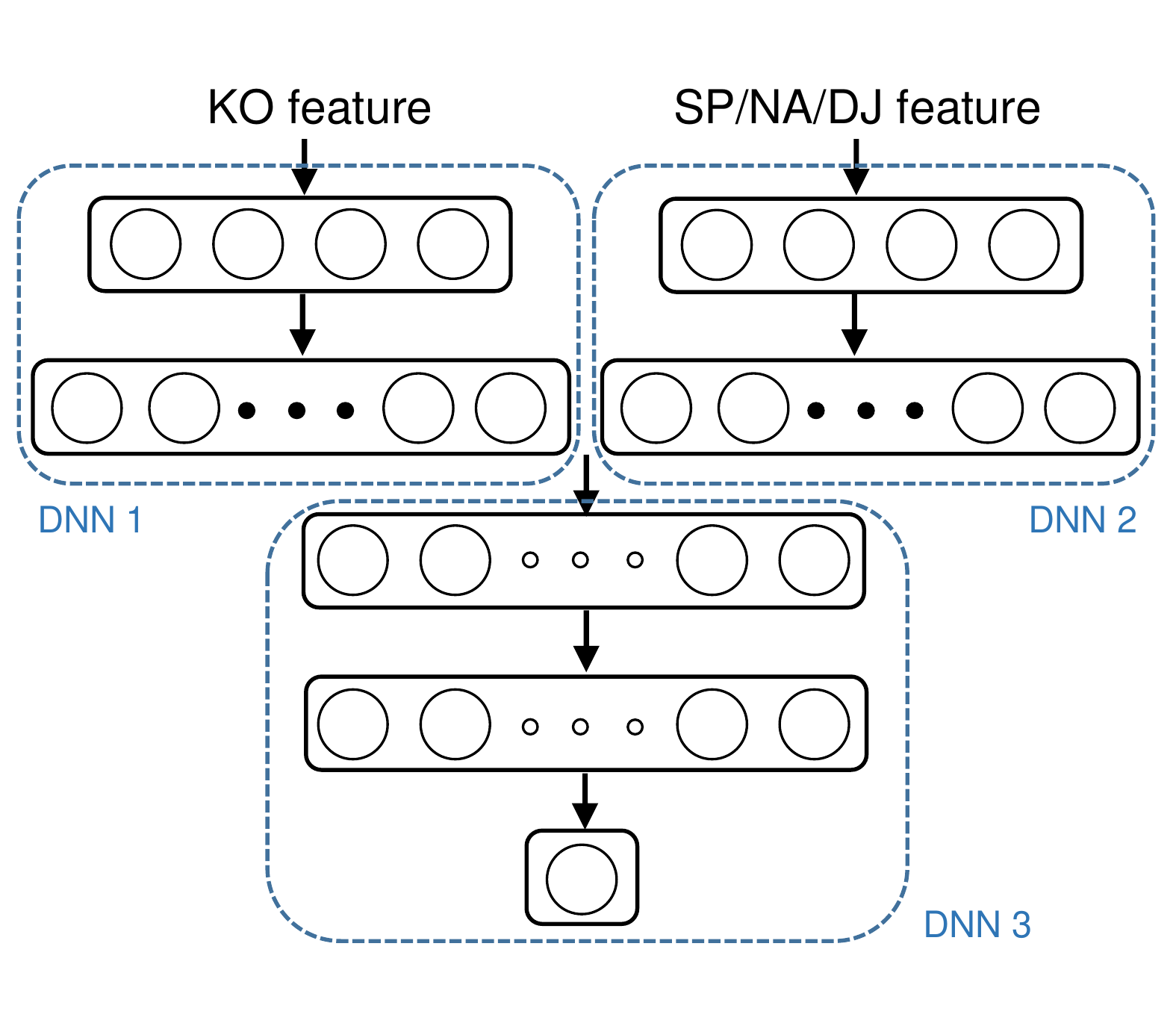}

   }
   }
\caption{The early fusion model is shown in the left figure, where the input feature is 
given by the concatenation of KO and US features. The intermediate fusion model is shown in the right figure, where the KO and US features are fed into DNN1 and DNN2 separately. The extracted features from these two DNNs are fused by DNN3 to generate
daily return predictions.}
\label{fig_DNN2}
\end{figure}

We built the prediction models based on
early, intermediate, and late fusion frameworks.
\\
\\
\noindent{\bf Single modal models (baseline)}:
To compare the performance of the fusion models, we used
four types of single modal models: (1) KO-Only DNN, 
(2) SP-Only DNN,
(3) NA-Only DNN, and
(4) DJ-Only DNN models (left-hand side of Fig. \ref{fig_DNN1}).
Their training sets $\{ {\boldsymbol x}_{t} \}$ are given by 
$\{ {\boldsymbol x}^{\textrm{KO}}_{t} \}$,
$\{ {\boldsymbol x}^{\textrm{SP}}_{t} \}$,
$\{ {\boldsymbol x}^{\textrm{NA}}_{t} \}$, and
$\{ {\boldsymbol x}^{\textrm{DJ}}_{t} \}$,
respectively.
\\
\\

\noindent {\bf Early fusion}: 
The input feature vectors 
are simply concatenated together at the input layer and then processed
together throughout the DNN 
(left-hand side of Fig. \ref{fig_DNN2}).
The feature vector is given by
\begin{align}
{\boldsymbol x}_{t}=[{\boldsymbol x}^{\textrm{KO}}_{t}; {\boldsymbol x}^{\textrm{US}}_{t}],
\end{align} 
where we use $[{\boldsymbol x}^{\textrm{KO}}_{t}; {\boldsymbol x}^{\textrm{US}}_{t}]$ to denote the concatenation of 
the two vectors ${\boldsymbol x}^{\textrm{KO}}_{t}$ and
${\boldsymbol x}^{\textrm{US}}_{t}$.
Although this model
is computationally efficient as compared to the other fusion models, as it requires a
lower number of parameters, it has several drawbacks
\cite{xu13} such as over-fitting in the case of a small-size training sample and the disregard of the specific 
statistical properties of each modality.
\\
\\
\noindent {\bf Intermediate fusion} 
Intermediate fusion combines the high-level features
learned by separate network branches (right-hand side of Fig. \ref{fig_DNN2}).
The network consists of two parts.
The first part consists of three independent deep neural networks, 
i.e., DNN1, which extracts features from the input feature $\{{\boldsymbol x}_{t}^{\textrm{KO}}\}$, 
DNN2, which extracts features from the input feature $\{{\boldsymbol x}_{t}^{\textrm{US}}\}$,
and DNN3, which fuses the extracted features 
and forecasts returns. The input feature vector
of DNN3 is given by
\begin{align}
{\boldsymbol a}^{\textrm{DNN3}}_{0}=
[{\boldsymbol a}^{\textrm{DNN1}}_{H};
{\boldsymbol a}^{\textrm{DNN2}}_{H}]
\end{align}
where 
${\boldsymbol a}^{\textrm{DNN1}}_{H}$ and
${\boldsymbol a}^{\textrm{DNN2}}_{H}$ are the units of
DNN1 and DNN2, respectively.
These fusions are not exposed to
cross-modality information at the raw data level and consequently, reveal
more intra-modality relationships than the early fusion model.
\\
\\
\noindent{{\bf Late Fusion}} 
Late fusion refers to the aggregation of decisions from 
multiple predictors (right-hand side of Fig. \ref{fig_DNN1}). 
Let $\hat{y}^{\textrm{KO}}_{t+1}$
and $\hat{y}^{\textrm{US}}_{t+1}$ 
be the predictions from the individual DNNs.
Then, the final prediction 
is
\begin{align}
\hat{r}_{t+1}=
F(\hat{r}^{\textrm{KO}}_{t+1}
, \hat{r}^{\textrm{US}}_{t+1}),
\end{align}
where $F$ is a rule combining the individual
predictions, such as 
 averaging \cite{shu16}, voting \cite{mor14}, or learned model \cite{glo11,ram11}, to generate the final results. In this study, we used
the linear rule given by
\begin{align}
\hat{r}_{t+1}=\lambda \times \hat{r}^{\textrm{KO}}_{t+1}
+(1-\lambda) \times \hat{r}^{\textrm{US}}_{t+1},
\end{align}
$\lambda$ is a weight to
combine the prediction values from the KO and US data.
Here, $\lambda$ is a mixing parameter 
that determines the relative contribution of each modality 
to the combined semantic space. We set 
$\lambda=0.5$, so that the KO and US sources contribute equally to the final prediction results.
\\
\\
All neural networks are trained by minimizing the mean squared error (MSE),
$(1/N)\sum_{t=1}^{N}(\hat{r}_{t+1}-r_{t+1})^{2}$, on the validation set.
 \\
 \\
\noindent {\bf Financial implications of fusion at different levels}

It would be financially meaningful to distinguish between fusion levels 
when international financial markets are combined.  
The price of domestic stock is commonly influenced by foreign events, but
the degree of that influence depends on the international
financial interdependency of the domestic stock market. 
Developed financial markets are likely to be 
highly exposed to international events and exhibiting high international correlations. In contrast,
underdeveloped markets are likely to be isolated and exhibiting low international and high intra-national correlations. 
Early fusion would be more suitable for 
developed markets in the sense that it can directly capture cross-correlations between domestic and foreign features in a single concatenation layer. In contrast, the intermediate fusion would be more suitable for underdeveloped
(or developing) markets in the sense that domestic features are more  
likely correlated with each other than with external foreign markets.

\section{Training}
\label{sec:5}
To find the best configuration, we used 
the tree-structured Parzen estimators (TPE) algorithm \cite{ber11}
, as one of the Bayesian hyper-parameter optimizations, which 
is capable of optimizing more hyperparameters simultaneously (Table \ref{params}). 
The hyper-parameters include: the number of layers, 
the number of hidden units per layer, the
activation function for a layer, the batch size, the optimizer, the 
learning rate, and the number of epochs. We apply the back-propagation algorithm 
\cite{wer74,rum86} to get the gradient of our models, without any pre-training, meaning that deep networks can be trained efficiently with ReLU without pre-training \cite{maa13}). All network weights were
initialized using Glorot normal initialization \cite{glo10}.
\\

\begin{table}[h]
\centering
\caption{List of hyperparameters and their corresponding range of
values.}
\label{table:meanerrorbaseline}
\begin{tabular}{lll}
\toprule
  Hyperparameter &\quad \quad \quad & Considered values/functions \\
\midrule
    Number of Hidden Layers && \{2, 3\} \\
    Number of Hidden Units && \{2, 4, 8, 16\} \\
    Dropout  && \{0.25, 0.5, 0.75\} \\
    Batch Size && \{32, 64, 128\} \\
    Optimizer && \{RMSProp, ADAM, SGD (no momentum)\} \\
    Activation Function&& Hidden layer: \{tanh, ReLU, sigmoid\}, Output layer: Linear \\
    Learning Rate && \{0.001\} \\
    Number of Epochs && \{100\} \\ 
\bottomrule
\end{tabular}
\parbox{\textwidth}{\small%
\vspace{1eX} 
{\bf Number of layers}: number of the layers of the (each branch) neural networks.
{\bf Number of hidden units}: number of units in the hidden layers
of the neural network.
{\bf Dropout}: dropout rates. 
{\bf Bath size}: number of samples per 
batch.  
{\bf Activation}: sigmoid function $\sigma(z)=1/(1+e^{-z})$, hyperbolic 
tangent function $\textrm{tanh}(z)=(e^{z}-e^{-z})/(e^{z}-e^{-z})$,
and rectified linear unit (ReLU) function $\textrm{ReLU}(z)=\textrm{max}(0,z)$.
{\bf Learning Rate}: learning rate of the back-propagation algorithm.
{\bf The Number of Epochs}: number of iterations over all the training data.
{\bf Optimizer}: stochastic gradient descent (SGD) \cite{kin14}, RMSProp \cite{ti12}, and ADAM \cite{kin14}
}
 \label{params}%
\end{table}

\subsection{Regularizations}
We used three types of regularization methods to control the overfitting of the networks and to improve the generalization error, including dropout, early stopping, and batch normalization. 
\\
\\
\noindent {\bf Dropout}. 
The basic idea behind dropout is to
temporarily remove a certain portion of hidden units from the network during training time, with the dropped units being randomly chosen at each and every iteration \cite{sr11}. This reduces the co-adaptation of the units, approximates model averaging, 
and provides a way to combine many different neural networks. In practice, dropout regularization requires specifying the dropout rates, which are the probabilities of dropping a neuron. 
In this study, we inserted dropout layers after every hidden layer, and performed a grid-search over the dropout rates of 0.25, 0.5, and 0.75 to find an optimal dropout rate for every architecture (Table \ref{params}).
\\
\\
{\bf Batch Normalization}
The basic idea of batch normalization (BN) is similar to that of
data normalization in training data pre-processing  \cite{se15}.
The BN technique uses the distribution of
the summed input to a neuron over a mini batch of training cases to compute the mean and variance, which are then used to normalize the summed input of that neuron on each training case. 
There is a lot of evidence that
the application of batch normalization results in
even faster convergence of training, increasing 
the accuracy compared to the same network without batch
normalization \cite{se15}.
\\
\\
\noindent {\bf Early Stopping}. Another approach we used to prevent overfitting is early stopping. 
Early stopping involves freezing the weights of neural networks at the epoch, where the validation error is minimal. The DNNs, which were trained with iterative back propagation, were able to learn the specific patterns of the training set after every epoch, instead of the general patterns, and begun to over-fit at a certain point. To avoid this problem, the DNNs were trained only with the training set, and the training was stopped if the validation MSE ceased to decrease for 10 epochs.

\section{Experiments}
\label{sec:6}
\subsection{Evaluation metric}
It is often observed that
the performance of stock prediction models
depends on the window size used. 
To make the evaluation task more robust,
we conducted experiments over three different windows: (1) Expt. 1 from 01-Jan-2006 to 31-Dec-2017;
Expt. 2 from 01-Jan-2010 to 31-Dec-2017; and
Expt. 3 from 01-Jan-2014 to 31-Dec-2017.

After obtaining the predictions for the test data, 
they were denormalized using the inverse formula of
Eq. (\ref{min-max}). Hereafter, $\hat{r}_{t}$ denotes
the denormalized prediction. 
Given a test set ${\{\boldsymbol x}^{\textrm{KO}}_{t}
,{\boldsymbol x}^{\textrm{US}}_{t}\}^{T}_{t=1}$ and a corresponding level
$\{ r_{t} \}^{T+1}_{t=2}$, where $T$ denotes the 
number of days in the test sample.
We evaluate the prediction performance using the MSE and the hit ratio defined as follows:
\begin{equation}
\textrm{Hit ratio}=\frac{1}{T}\sum_{t=1}^{T} P_{t},
\end{equation}
where $P_{t}$ is the directional movement of the prediction on the $t^{th}$ trading day, defined as:
\[ P_{t} =
  \begin{cases}
    1       & \quad \text{if   } \hat{r}_{t+1} \cdot \textrm{r}_{t+1}>0
 \textrm{\hspace{0.2cm}} \textrm{(i.e., correct directional prediction)}    
    , \\
    0  & \quad  \text{otherwise}  \textrm{\hspace{1.2cm}} \textrm{(i.e., incorrect directional prediction)}.\\
  \end{cases}
\]

\subsection{Daily strategies as baselines}
To evaluate the single and fusion models, we examined
the hit ratios for the three regular rules:
\begin{itemize}
\item {\bf Momentum-based prediction-\RNum{1}}:
If the KOSPI index rises (falls) today, 
it predicts that the KOSPI index will rise (fall) tomorrow too.
\item {\bf Momentum-based prediction-\RNum{2}}: 
If the S$\&$P500 index rises (falls) today, it predicts that the KOSPI index will rise (fall) tomorrow too.
\item {\bf Buy and holding strategy}: 
Based on positive historical returns, it predicts that the KOSPI index of the next day will rise.
\end{itemize}
\begin{table}[htbp]
  \centering
  \caption{Hit ratios of the three regular rules.}
    \begin{tabular}{ccccccc}
    \toprule
    Expt. no      & $\quad$ & Momentum-based prediction-\RNum{1} &$\quad$ & Momentum-based prediction-\RNum{2} &$\quad$& Buy and hold \\
    \midrule
    1    && 0.484  && {\bf 0.562} &&0.549 \\
    2    && 0.492  && {\bf 0.558} &&0.534 \\
    3    && 0.488  && {\bf 0.536} &&0.523\\
    \bottomrule
    \end{tabular}%
  \label{baseline}%
\end{table}%
Table \ref{baseline} shows that the momentum-based prediction-\RNum{2} is the most accurate of the three rules, exhibiting hit ratios of 0.562, 0.558, and 0.536 for Expt. 1, Expt. 2, and Expt. 3, respectively.

\subsection{Results}
We present the prediction results obtained with the fusion models
for the pairs of KO and SP features (Table. \ref{MultimodalAccuracy1}),
the KO and NA features (Table \ref{MultimodalAccuracy2}), 
and the KO and DJ features (Table \ref{MultimodalAccuracy3}),
along with those of non-fusion models for each of
the country-specific features as a baseline model. To remove potentially undesirable variances that arise from parameters having different min-max ranges, 
we also conducted the experiments with three distinct ranges $[-1, 1]$, $[0,1]$, 
and $[-0.5,0.5]$. 
The main findings are follows:
\\
\\
\noindent {\bf Early vs. intermediate fusion}
For the fusion of the KO and SP features (Table \ref{MultimodalAccuracy1}), 
the mean hit ratio (directional prediction) of the early fusion ($0.606\pm0.011$) is slightly higher or
comparable to that of the intermediate fusion ($0.597\pm 0.029$).
For the fusion of the KO and NA features (Table \ref{MultimodalAccuracy2}), 
the hit ratio of early fusion ($0.584 \pm 0.019$) is 
slightly lower or
comparable to that of the intermediate fusion ($0.595\pm 0.013$).
For the fusion of the KO and DJ features (Table \ref{MultimodalAccuracy3}), 
the hit ratio of the early fusion ($0.585\pm0.027$) 
is slightly lower or
comparable to that of the intermediate fusion ($0.600\pm0.022$). 
Thus, the performance of the two fusion approaches are comparable overall, which is consistent over the different window sizes and the min-max ranges.
In terms of computational efficiency, the early fusion model is more attractive due to its lower number of parameters compared to the intermediate fusion model.
\\
\\
\noindent {\bf Single vs. multimodality}
The overall hit ratio of the single modal models is about 0.49:  0.499$\pm$0.028 
for the KO-Only DNN,
0.496$\pm$0.023 for the SP-Only DNN (Table \ref{MultimodalAccuracy1}),
0.497$\pm$0.024 for the NA-Only DNN
(Table \ref{MultimodalAccuracy2}), and
0.492$\pm$0.029 for the DJ-only DNN (Table \ref{MultimodalAccuracy3}).
Interestingly, the performances are
slightly worse than the momentum-based prediction-\RNum{2} (approximately 0.55) and the buy and hold strategy (approximately 0.53). The hit ratios of the late fusion 
are $0.499\pm0.024$ for the KO and SP fusion, $0.498\pm0.021$ for the KO and NA fusion,
and $0.496 \pm 0.026$ for the KO and DJ fusion, which are lower than those of the early and intermediate fusions.
These results show that the parameters of the two modalities 
needs to be estimated jointly. The poor performance of the single modality models clearly emphasizes the importance of multimodal integration 
to leverage the complementarity of stock data
and provide more robust predictions.

\begin{table}[htbp]
  \centering
  \caption{Hit ratio (MSE$\times 10^{-5}$) measure for Expts. 1-3 for the KO and SP data}
    \begin{tabular}{ccccccccccc}
    \toprule
    \multirow{3}[0]{*}{Scaling} & &\multicolumn{3}{c}{Non-fusion} & \multicolumn{6}{c}{Multimodal fusion}
     \\
     \cmidrule(lr){3-6}
     \cmidrule(lr){7-11}
           &   & KO-only DNN   &  & SP-only DNN & & Late &  & Early &   & Intermediate \\

    \midrule
        & & \multicolumn{9}{c}{Expt. 1} \\
                    \cmidrule{3-11}\morecmidrules\cmidrule{3-11}
     $[-1,1]$    & &0.490 (5.257)&&0.499 (5.268)& &0.513 (5.26) & &  {\bf 0.609} ({\bf 4.781})&&  0.599 (4.989)\\
      $[0,1]$    & &0.526 (5.284)&&0.506 (8.787)& &0.514 (6.011) & & {\bf 0.612} ({\bf 4.630}) && 0.592 (0.480) \\
      $[-0.5,0.5]$& &0.519 (5.622)&&0.500 (7.590)& &0.501 (5.851) & & 0.607 ({\bf 0.463})       && {\bf 0.608} (4.820) \\
        & & \multicolumn{9}{c}{Expt. 2} \\
                    \cmidrule{3-11}\morecmidrules\cmidrule{3-11}
     $[-1,1]$  &   &0.505 (5.726)&&4.755 (6.199)& &0.479 (5.852) &  & 0.613 ({\bf 4.951})    &&  {\bf 0.617} (5.091)\\
     $[0,1]$    &  &0.487 (5.717)&&4.755 (6.343)&  &0.470 (5.917)& & {\bf 0.615} ({\bf 5.054}) &&  0.587 (5.193)\\
     $[-0.5,0.5]$& &0.484 (5.716)&&4.755 (6.437)&  &0.477 (5.933) & & 0.590 ({\bf 4.982})&&{\bf 0.648} (5.048)\\
        & & \multicolumn{9}{c}{Expt. 3} \\
                    \cmidrule{3-11}\morecmidrules\cmidrule{3-11}
     $[-1,1]$& & 0.552 (3.895)&&0.549 (3.902)& &0.549 (3.891) & &0.602 ({\bf 3.601})&&{\bf 0.609} (3.629)\\
     $[0,1]$ & & 0.464 (4.004)& &0.507 (4.558)  & &0.500 (4.132)& & {\bf 0.619} ({\bf 3.680})  &    &  0.545 (3.890)\\
    $[-0.5,0.5]$& & 0.468 (3.991)&  & 0.482 (4.877)  & &0.496 (4.251) &  & {\bf 0.584} ({\bf 3.676}) &    &  0.570 (3.700)\\
      \midrule
         Mean$\pm$SD    & &0.499$\pm$0.028&&0.496 $\pm$0.023& &0.499$\pm$ 0.024& & {\bf 0.606}$\pm$0.011      && 0.597$\pm$0.029 \\
    \bottomrule
    \end{tabular}%
  \label{MultimodalAccuracy1}%
\end{table}%

\begin{table}[htbp]
  \centering
  \caption{Hit ratio (MSE$\times 10^{-5}$) measure for Expts. 1-3 for the KO and NA data}
    \begin{tabular}{ccccccccccc}
    \toprule
    \multirow{3}[0]{*}{Scaling} & &\multicolumn{3}{c}{Non-fusion} & \multicolumn{6}{c}{Multimodal fusion}
     \\
     \cmidrule(lr){3-6}
     \cmidrule(lr){7-11}
            &  & KO-Only DNN   &  & NA-Only DNN & &
            Late&   & Early &   & Intermediate \\
    \midrule
        &  & \multicolumn{9}{c}{Expt. 1} \\
                    \cmidrule{3-11}\morecmidrules\cmidrule{3-11}
     $[-1,1]$   & & 0.490 (5.257)& & 0.518 (5.290)  & &0.500 (5.275) & &  0.598 (4.774) &    &  {\bf 0.600} ({\bf 4.586})\\
      $[0,1]$    & & 0.526 (5.284)& &0.500 (6.699)   & &0.504 (5.571) &  & 0.594 (4.479) &     & {\bf 0.596} ({\bf 0.478}) \\
      $[-0.5,0.5]$ & & 0.519 (5.262)& & 0.509 (9.565)   & &0.508 (6.351) &  & 0.592 ({\bf 0.467}) &     & {\bf 0.605} (5.048) \\
          && \multicolumn{9}{c}{Expt. 2} \\
                    \cmidrule{3-11}\morecmidrules\cmidrule{3-11}
     $[-1,1]$   & & 0.505 (5.726)&& 0.484 (6.090) &   &0.482 (5.823) & & {\bf 0.608} (5.218)    &  &  0.597 ({\bf 5.178})\\
      $[0,1]$    & & 0.487 (5.717)&&  0.480 (6.061)  & &0.486 (5.826)&  & 0.597 ({\bf 5.071}) &     & {\bf 0.613} (5.122)\\
      $[-0.5,0.5]$ & & 0.484 (5.716)&& 0.475 (6.162)   & &0.475 (5.841) &  & {\bf 0.580} ({\bf 5.196})  &    & 0.573 (5.221)\\
          && \multicolumn{9}{c}{Expt. 3} \\
                    \cmidrule{3-11}\morecmidrules\cmidrule{3-11}
     $[-1,1]$ &  &0.552 (3.895)& &0.549 (3.908) & &0.549 (3.901) & & 0.556 ({\bf 3.624})   &  &  {\bf 0.605} (3.690)\\
     $[0,1]$  & &0.464 (4.004) && 0.471 (4.521)   & &0.496 (4.962) & &  0.552 (3.710)  &    &  {\bf 0.57}3 ({\bf 3.693})\\
    $[-0.5,0.5]$ & & 0.468 (3.991) & & 0.489 (4.810) & &0.489 (4.244) &   & 0.584 ({\bf 3.648}) & & {\bf  0.599} (3.764)\\
                  \midrule
       Mean$\pm$SD     & &0.499$\pm$0.028&&0.497 $\pm$0.024& &0.498$\pm$0.021& & 0.584$\pm$ 0.019      && {\bf 0.595}$\pm$0.013 \\
    \bottomrule
    \end{tabular}%
  \label{MultimodalAccuracy2}%
\end{table}%

\begin{table}[htbp]
  \centering
  \caption{Hit ratio (MSE$\times 10^{-5}$) measure for Expts. 1-3 for the KO and DJ data}
    \begin{tabular}{ccccccccccc}
    \toprule
    \multirow{3}[0]{*}{Scaling} & &\multicolumn{3}{c}{Non-fusion} & \multicolumn{6}{c}{Multimodal fusion}
     \\
     \cmidrule(lr){3-6}
     \cmidrule(lr){7-11}
             &  & KO-Only DNN   &  & DJ-Only DNN & &Late&   & Early &   & Intermediate  \\

         \midrule
         & & \multicolumn{9}{c}{Expt. 1} \\
                    \cmidrule{3-11}\morecmidrules\cmidrule{3-11}
     $[-1,1]$   & &0.490 (5.257) & & 0.518 (5.263)  & &0.515 (5.253)& &  0.598 ({\bf 4.774}) &    &  {\bf 0.623} (4.987)\\
      $[0,1]$   & &0.526 (5.284)  & & 0.521 (10.360)  & &0.523 (7.186)&  & 0.607 ({\bf 4.861}) &     & {\bf 0.617} (5.539) \\
      $[-0.5,0.5]$ && 0.519 (5.262) & &  0.483 (0.701)  & &0.488 (5.698)&  & {\bf 0.610} ({\bf 0.517}) &     & 0.609 (5.748) \\
        &  & \multicolumn{9}{c}{Expt. 2} \\
                    \cmidrule{3-11}\morecmidrules\cmidrule{3-11}
     $[-1,1]$  & &0.505 (5.726) && 0.473 (6.241)  &  &0.473 (5.861)&  & 0.603 ({\bf 5.219})    &  &  {\bf 0.606} (5.287)\\
      $[0,1]$  &  & 0.487 (5.717) && 0.475 (6.076)   & &0.482 (0.582)&  & 0.601 (5.161) &     & {\bf 0.613} ({\bf 5.122})\\
      $[-0.5,0.5]$ && 0.484 (5.716)&  & 0.473 (6.465)  & &0.472 (5.892)&  & 0.592 ({\bf 5.326})  &    & {\bf 0.617} (5.694)\\
        &  & \multicolumn{9}{c}{Expt. 3} \\
                    \cmidrule{3-11}\morecmidrules\cmidrule{3-11}
     $[-1,1]$ &  & 0.552 (3.895)&& 0.549 (3.906) & &0.549 (3.902)& & 0.577 ({\bf 3.825})   &  &  {\bf 0.580} (4.015)\\
     $[0,1]$  & &0.464 (4.004) && 0.482 (4.514)   & &0.482 (4.093)& &  0.538 (3.933)  &    &  {\bf 0.556} ({\bf 3.890})\\
    $[-0.5,0.5]$ && 0.468 (3.991) && 0.461 (4.693)  & &0.482 (4.171)&   & 0.542 ({\bf 3.878}) & & {\bf  0.584} (4.015)\\
                        \midrule
       Mean$\pm$SD    & &0.499$\pm$0.028&&0.492 
       $\pm$0.029& &0.496$\pm$0.026 & & 0.585$\pm$0.027      && {\bf 0.600}$\pm$0.022 \\
    \bottomrule
    \end{tabular}%
  \label{MultimodalAccuracy3}%
\end{table}%

\section{Discussion and conclusion}\label{sec:7}
We developed stock prediction models that combine information from the South Korean and US stock markets by using multimodal deep learning. 
We exploited DNN as a branch of deep learning to take advantage of
its strong capability in non-linear modeling and designed three types of architectures to capture the cross-modal correlation at different levels.
Experimental results show that the early and intermediate fusion models 
predict stock returns more accurately than the single modal and late fusion models, which do not consider cross-modal correlation in their predictions. 
This indicates that joint optimization can effectively capture complementary information between the markets and assist in the improvement of stock predictions.

This study has a few limitations. 
First, we examined three different time periods of $2006-2017$, $2010-2017$, and $2014-2017$.
Over these periods, the early and intermediate fusion model consistently outperformed the regular rule-based prediction and late fusion models, in terms of accuracy. However, the sample sizes of the present study are relatively small, and the performance of the models may vary based on the period and depending on the globalization level of the stock markets. Second,
the information of international markets is limited to trading data.

Future works will focus on two aspects. First, we plan to include more diverse information sources such as fundamental data and
sentiment indexes. The stock prices are determined by the supply and demand of the stocks, which occurs due to various information inputs. Thus, integrating more diverse data would lead to an improvement in the reliability of stock prediction.
Second, we plan to analyze the prediction results by using 
explainable machine learning techniques.
Understanding and interpreting prediction models are crucial in financial fields.  

\subsubsection*{Acknowledgments} 
This work was supported by the ICT R$\&$D program of MSIP/IITP [2017-0-00302, Development of Self Evolutionary AI Investing Technology].

\end{document}